%% file: pca_dibs.tex
\documentclass[twocolumn]{aastex61}

\usepackage{amsmath}
\usepackage{graphicx}

\shorttitle{A PCA of the DIBs}
\shortauthors{Ensor et al.}

\let\oldhat\hat
\renewcommand{\hat}[1]{\oldhat{\mathbf{#1}}}

\begin{document}

\title{A Principal component analysis of the diffuse interstellar bands}

\author{T. Ensor}
\affil{Department of Physics and Astronomy and Centre for Planetary
  Science and Exploration (CPSX), The University of Western
  Ontario, London, ON N6A 3K7, Canada}
\author[0000-0002-2666-9234]{J. Cami}
\correspondingauthor{J. Cami}
\affil{Department of Physics and Astronomy and Centre for Planetary
  Science and Exploration (CPSX), The University of Western
  Ontario, London, ON N6A 3K7, Canada}
\affil{SETI Institute, 189 Bernardo Ave, Suite 100, Mountain View, CA
  94043, USA} 
\author{N.H. Bhatt}
\affil{Department of Physics and Astronomy and Centre for Planetary
  Science and Exploration (CPSX), The University of Western
  Ontario, London, ON N6A 3K7, Canada}
\author{A. Soddu}
\affil{Department of Physics and Astronomy and Centre for Planetary
  Science and Exploration (CPSX), The University of Western
  Ontario, London, ON N6A 3K7, Canada}

\begin{abstract}
We present a principal component analysis of 23 line of sight
parameters (including the strengths of 16 diffuse interstellar bands,
DIBs) for a well-chosen sample of single-cloud sightlines representing
a broad range of environmental conditions. Our analysis indicates that
the majority ($\sim$93\%) of the variations in the measurements can be
captured by only four parameters The main driver (i.e., the first
principal component) is the amount of DIB-producing material in the
line of sight, a quantity that is extremely well traced by the
equivalent width of the $\lambda$5797 DIB. The second principal
component is the amount of UV radiation, which correlates well with
the $\lambda$5797/$\lambda$5780 DIB strength ratio. The remaining two
principal components are more difficult to interpret, but are likely
related to the properties of dust in the line of sight (e.g., the
gas-to-dust ratio). With our PCA results, the DIBs can then be used to
estimate these line of sight parameters.
\end{abstract}

\keywords{ISM:lines and bands --- ISM:molecules --- methods:data
  analysis --- methods:statistical}

\section{Introduction} \label{sec:intro}
One of the greatest outstanding astronomical challenges is the
identification of the diffuse interstellar bands (DIBs): a series of
$\sim$500 absorption features detected in optical and infrared spectra
toward reddened stars (see \citealt{Herbig1995,Sarre2006,Snow:DIBconf}
for reviews and \citealt{Hobbs:204827,Hobbs:183143} for recent
surveys). It has been clear that the DIBs arise from material in
interstellar clouds since they were first detected \citep{Heger1922};
but despite nearly 100 years of research, most of the DIB carriers
remain unidentified. The only notable exception is the recent
identification of four DIBs as due to C$_{60}^+$ \citep{C60_1,
  C60_2}. This identification is in line with the general consensus
that the DIB carriers are highly stable, carbonaceous, gas-phase
molecules.

An identification of DIBs with specific carriers requires a perfect
match between laboratory spectra and astronomical observations;
however, given the countless numbers of possible carrier candidates,
this is not an easy task. To guide these laboratory efforts,
observational studies aim to learn about the nature of the carriers
and constrain the set of possible species. Two types of such studies
that are particularly relevant for this paper are correlation studies
-- either mutual DIB correlations, or correlations between the DIBs
and line of sight properties \citep[see
  e.g.][]{Seab1984, Herbig1993, Cami1997, McCall2010, Friedman2011} --
and research into the environmental behavior of the DIBs
\citep[e.g.][]{Jenniskens:OrionDIBs, Cami1997, Sonnentrucker1997,
  Cox:LMC-DIBs}.

The basic idea behind pairwise correlations is simple: If two DIBs
arise from the same state in the same carrier, they should have the
same strength ratio in all lines of sight and thus, their equivalent
widths (EWs) should exhibit a perfect correlation. Observational
studies have not found two DIBs that show such a perfect
correlation. The best case is the $\lambda$6196 and $\lambda$6614 DIBs
which correlate well in a large sample of sightlines
\citep[correlation coefficient $r$ of 0.986; see][]{McCall2010}.
However, even these two DIBs show quite different behavior in the
remarkable sightline towards Herschel 36 implying that they are most
likely originating from different carriers
\citep{Dahlstrom:Herschel36,Oka:Herschel36}. This has led to the ``one
DIB, one carrier'' paradigm \citep{Herbig1995, Cami1997,
  Snow:DIBconf}. At the same time, the notion of DIB ``families'' can
be established: sets of DIBs that correlate fairly well with one
another and that might have similar or chemically related carriers
\citep{3families, Cami1997}. There are two important caveats though in
correlation studies. First, correlation studies generally include only
a small number of DIBs, typically fairly strong and narrow
DIBs. Second, while the role of measurement uncertainties on the
correlation coefficient is well established \citep[see
  e.g. discussions in][]{Herbig:1975, Cami1997}, they are often not
taken into account in correlation studies.

Correlations between the DIB strengths and line of sight parameters
can reveal additional properties about the DIB carriers. DIB strengths
often show some correlation ($r$ typically 0.7) with various other
line of sight parameters, albeit typically with a large scatter around
the mean relation; examples are the correlation with $E(B-V)$, or the
$\lambda$5780 DIB strength with N(\ion{H}{1}) or the column densities
of other interstellar species \citep[see e.g.][]{Herbig:1975,
  Herbig1993, Herbig1995, Krel1999, Welty2006, Friedman2011,
  Lan:SDSS_DIBs, Baron:SDSS}. A particularly intriguing finding is a
subset of DIBs (the so-called ``C$_2$-DIBs'') that roughly correlate
with N(C$_2$) and are thought to be chemically related to C$_2$ or
else form under similar conditions \citep{Thorburn}. Modern surveys
have confirmed such relations for averaged DIB strengths on large
scales, and have furthermore also shown that much of the scatter can
be traced back to differences in the amounts of H$_2$ relative to
\ion{H}{1} in the line of sight \citep{Herbig1993, Lan:SDSS_DIBs}.

Part of the scatter in these correlations must thus be due to changes
in the physical environment that drive the carrier abundances. Indeed,
the DIBs exhibit clear environmental behavior, and show intensity
variations that could be explained for instance by ionization or
\mbox{(de-)}hydrogenation \citep{Jenniskens:OrionDIBs, Cami1997,
  Sonnentrucker1997}. Interestingly, various parameters have been
shown to be indicative of these environmental conditions. The strength
ratio between two strong DIBs, $\lambda$5797 and $\lambda$5780, is
highly variable and a good indicator of local conditions
\citep{Krelowski:MoreDibFamilies}. Using this ratio, diffuse clouds
can typically be subcategorized into two groups -- $\sigma$ and
$\zeta$ type clouds, named after their prototypes $\sigma$~Sco
(HD~147165) and $\zeta$~Oph (HD~149757). $\sigma$ clouds have lower
W($\lambda$5797)/W($\lambda$5780) ratios and are characterized by stronger
UV exposure; $\zeta$ clouds, on the other hand, probe deeper layers of
diffuse clouds where material is sheltered from UV radiation, and this
causes a much larger W($\lambda$5797)/W($\lambda$5780) ratio while
simultaneously affecting the dust properties \citep{Cami1997}.

The large number of DIBs coupled with the lack of strong correlations
suggest that the DIBs carry an enormous diagnostic potential to study
the environments in which they reside. At the same time, it raises the
question of what factors drive these variations in the DIB strengths,
and how it is possible that there is such a lack of correlations in
such a large collection of spectral lines. Here, we address these
questions, and in particular the key question: {\em how many
  parameters do we need to explain the variations in the DIB spectrum
  and what are those parameters?}
    
To this end, we present a multivariate analysis of a set of strong and
clean DIBs with several line of sight parameters. In a
proof-of-concept study, we first perform a principal component
analysis (PCA) on the data to find out how many parameters are
required to describe the observed variations among the DIBs. We
physically interpret these new parameters and find convenient
quantitative alternatives to represent these parameters. From this
work, the huge diagnostic potential of the DIBs becomes clear: since
DIBs are products of their environments, we can use DIBs to determine
physical parameters of their environment -- even without identifying
the carriers.

This paper is organized as follows. In Section~\ref{sec:Methods}, we
describe our sample selection and how we acquired our data. We then
describe PCA in Section~\ref{sec:PCA}, followed by our results in
Section~\ref{sec:full_PCA}. We interpret the results in
Section~\ref{sec:discussion} and present our conclusions in
Section~\ref{sec:Conclusions}.

\section{Data, Observations \& Methods}
\label{sec:Methods}

\subsection{Target Selection}
\label{subsec:targets}

Our goal in this paper is to determine the parameters that drive the
variations in the DIB spectrum. Thus, we need to select a sample of
sightlines where physical conditions are reasonably well determined,
and that represent the overall observed variations of the DIB spectrum. The
first requirement implies that we should restrict ourselves as much as
possible to single-cloud lines of sight to avoid having to deal with
ill-defined averages throughout multiple intervening clouds. The
second requirement stresses the importance of including lines of sight
that are as observationally different as possible. Finally, in order
to be able to relate any changes to known observables, we need to pick
lines of sight for which auxiliary data (e.g. hydrogen column
densities, $E(B-V)$, extinction properties, \dots) are known and
available. In this pilot study, we chose to restrict ourselves to
include only a limited number of DIBs, and to lines of sight for which
we can find high-resolution spectra that allow us to exclude possible
blends with stellar lines. 

We started our selection of targets from the thorough and detailed
study of elemental depletion in the lines of sight towards 243 stars
published by \citet{Jenkins}. This study critically reviews available
literature data for E(B-V), N(\ion{H}{1}), N(H$_2$), N(H), and
introduces a depletion strength factor, F$_\star$, describing the
collective level of elemental depletion in a line of sight. Starting
from this sample thus ensures a consistent treatment of the required
auxiliary line-of-sight data and allows us to ensure a wide coverage
of environmental conditions (to the extent that they can be traced by
any of these parameters).

We searched the VLT/UVES and ELODIE archives to find good-quality,
high-resolution spectra of these targets. UVES (the Ultraviolet and
Visual Echelle Spectrograph) is a high-resolution instrument on the
VLT covering wavelength ranges from 3000 - 4000\AA\ and 4200 -
11,000\AA, with maximum resolutions of 80,000 and 110,000,
respectively \citep{UVES}. ELODIE is an echelle spectrograph on the
1.93m telescope at the Observatoire de Haute-Provence in
France. ELODIE has a resolution of 42,000 and covers the wavelength
range from 3906 - 6801\AA\ \citep{Elodie}. We found appropriate data
for 91 of the Jenkins targets: 43 targets in the UVES database; the
remaining 48 from the ELODIE database. In a few rare cases, parts of
the spectrum would be of too low quality, or simply missing from the
data, and in those cases we supplemented our data with archival
spectra from the ESPaDOns (Echelle spectropolarimetric device for the
observation of stars) instrument on the Canada-France-Hawaii Telescope
(CFHT) -- a bimodal instrument with maximum resolutions of 68,000 and
81,000 for its spectropolarimetric and non-polarimetric modes,
respectively, covering a wavelength range of 3700 -
10,000\AA\ \citep{Espadons}.

\begin{figure}
\centering
\resizebox{8cm}{!}{\includegraphics[trim=2cm 7cm 3cm 6cm]{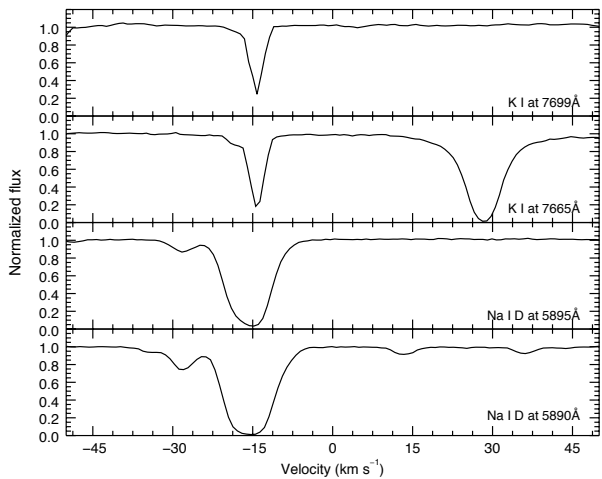}}
\caption{\label{fig:Singlecloud}Interstellar lines for one of our
  targets, HD~149757 in the heliocentric rest frame. Here, the
  \ion{Na}{1} lines show a dominant component at approximately
  -15~km~s$^{-1}$, as well as a second component that is much weaker
  near -27~km~s$^{-1}$. The \ion{Na}{1} lines are somewhat saturated,
  but the \ion{K}{1} lines confirm that there is just a single,
  dominant radial velocity component. This star meets our criterion
  for a single-cloud line of sight. Note that the feature near
  +27~km~s$^{-1}$ in the \ion{K}{1} (7665~\AA) plot is a telluric
  oxygen line and not of interstellar origin.}
\end{figure}

To further select only single-cloud lines of sight, we examined the
interstellar \ion{Na}{1} D lines at 5890 and 5895\AA\ and the
\ion{K}{1} lines at 7665 and 7699\AA\ (see e.g., \cite{Bhatt2015},
illustrated in Figure~\ref{fig:Singlecloud}). For our current study,
we consider a sightline to be a single cloud if these interstellar
lines show only one dominant component at the spectral resolution of
UVES or ELODIE. Thus, a target was still considered to be a single
cloud if there were multiple radial velocity components, but one
component had significantly stronger features than the others. For
example, HD~149757 is known to have two strong radial velocity
components at approximately $-27$ and $-15$~km~s$^{-1}$, but the one
at $-15$~km~s$^{-1}$ has much larger column densities \citep[][see
  also Fig.~\ref{fig:Singlecloud}]{Herbig_zetaOph}. For the targets
with only ELODIE spectra, the \ion{K}{1} lines are outside the
available wavelength range, and thus we could only use the \ion{Na}{1}
D lines to inspect the number of radial velocity components. An
obvious and inherent drawback of using these lines is that they are
easily saturated. We searched for \ion{Ca}{1} and CH lines too, but in
most cases, these were too weak to be seen.

With this exercise, we established that 33 of our targets can be
identified as single-cloud lines of sight. However, for three of them,
the UVES archival spectra have a gap in the wavelength coverage from
approximately 5760-5830\AA\ and thus two important DIBs,
$\lambda$5780 and $\lambda$5797 are missing. We therefore excluded
these targets from our data set. After these considerations, we were
left with a sample of 30 single-cloud lines of sight. These targets
are listed in Table~\ref{table:target_info} along with their line of
sight parameters we will use in this paper (see below). 

It should be noted that six of the targets in our sample -- HD~23630,
HD~24534, HD~110432, HD~149757, HD~164284, and HD~202904 -- are Be
stars. Such objects have an intrinsic E(B-V) value relative to non-Be
stars of the same spectral type \citep{Schild_Be_stars,
  Sigut_Be_stars}; the hot, circumstellar gas produces excess emission
in the V filter and therefore, their E(B-V) values are
over-estimated. Furthermore, any dust existing in the circumstellar
shell (CS) can contribute to E(B-V), while there is no evidence to
suggest that DIBs exist in CS environments \citep{Krelowski&Sneden,
  Snow_CS_1,Snow_CS_2}. Hence, we expect weaker-than-normal DIB
strengths relative to E(B-V) for these six targets.

We applied a heliocentric correction to all targets, and then shifted
all spectra to their interstellar rest frames. Interstellar velocity
components were obtained from the literature, or measured from a known
interstellar feature, if not available. These velocities are listed in
column 14 of Table~\ref{table:target_info}.

\subsection{DIB Measurements}
\label{subsec:measurements}

For our purposes, we only wanted to include those DIBs whose
equivalent widths can be confidently measured, i.e., with small
relative errors. This limits our selection to fairly strong and often
narrow DIBs that are as much as possible free of contamination from
stellar lines. The only exceptions we made was to include four of the
C$_2$ DIBs -- $\lambda\lambda$4964, 5513, 5546, and 5769 -- despite
the latter three being quite weak. We thus include a sample of 16 DIBs
in our analysis: $\lambda\lambda$4428, 4964, 5494, 5513, 5545, 5546,
5769, 5780, 5797, 5850, 6196, 6270, 6284, 6376, 6379, and 6614. 

\begin{figure}
\centering
\resizebox{8.5cm}{!}{\includegraphics[trim={2cm 10.5cm 2cm 10.5cm}]{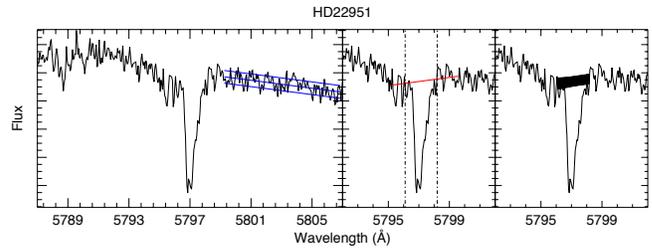}}
\caption{\label{fig:ew_measurement}Illustration of measuring the
  equivalent width of the $\lambda$5797 DIB for the target
  HD~22951. (\textit{Left:}) In a featureless part of the spectrum
  near the feature, we measured the S/N by determining the standard
  deviation of flux values across a straight line. The blue parallel
  lines show the mean flux value and the $\pm 1\sigma$
  values. (\textit{Center:}) We chose a point on either side of the
  feature, and determined a linear continuum between the two (shown in
  red). The chosen integration limits are shown as dotted
  lines. (\textit{Right:}) This figure shows the 1,000 instances of
  the continuum slopes generated through our Monte Carlo simulation,
  superimposed over the spectrum. }
\end{figure}

We measured the equivalent widths for these 16 DIBs in all lines of
sight. Most of these were straightforward to measure, as they are not
heavily contaminated by stellar and/or telluric features. The largest
uncertainty on these measurement stems from establishing the
continuum. To obtain a good estimate of these uncertainties, we used
the following Monte Carlo approach to vary the continuum level and
perform direct integration of the spectra (similar to that in
\cite{Bhatt2015}; see Figure~\ref{fig:ew_measurement} for
illustration). First, we measured the standard deviation of the flux
values over a specified, featureless range of data in the vicinity of
the feature to estimate the uncertainty on the flux values -- i.e. we
measured the signal-to-noise ratio (S/N). Next, we selected a point on
each side of the feature and defined a continuum baseline by adopting
a linear continuum between those two points. We also selected points
as our integration limits; note that for consistency, we used the same
integration limits for the same feature in all lines of sight. To
simulate the process of determining the continuum, we varied the two
selected continuum points 1,000 times by adding a random number to the
flux values selected from a normal distribution with a mean of zero
and standard deviation equal to $\frac{1}{3}$ the measured standard
deviation in the featureless continuum; we believe that this
represents well how accurately one can position the continuum (which
corresponds to determining the mean flux in the adjacent continuum),
and we found that this produces reasonable continuum estimates (see
Figure~\ref{fig:ew_measurement}). Using one full standard deviation
produces many continuum points which are clearly too high or too low
and thus result in unrealistic continuum levels. In essence, we thus
simulated the entire process of determining a continuum line 1,000
times. For each continuum, we then measured the equivalent width. The
equivalent width we use in this paper is then the mean of these 1,000
measurements, and the standard deviation of these measurements
provides the uncertainty.

We kept a few precautions in mind when using this method. For
instance, the sharp and narrow $\lambda$5797 is known to be blended
with the broader and shallower $\lambda$5795. To avoid measuring a
contribution from $\lambda$5795, we measured the $\lambda$5797 while
treating the $\lambda$5795 as continuum (see
Figure~\ref{fig:ew_measurement}). If bad pixels were found within the
integration range, that data point was replaced with the average of
the neighboring points. For HD~23180, the $\lambda$5494 feature was
strongly contaminated by a stellar line. We could not resolve the two
features to obtain a proper measurement, so instead we adopted the
value obtained from higher resolution observations by \cite{Bondar}.

The two broad DIBs in our sample -- $\lambda$4428 and $\lambda$6284 --
cannot be measured using the methods described above, because they are
heavily contaminated by stellar and telluric features,
respectively. For the $\lambda$6284 DIB, we first applied a telluric
correction using \verb+molecfit+ (version 1.1.0) \citep{Molecfit1,
  Molecfit2} and then proceeded as for the other DIBs.

The $\lambda$4428 DIB is extremely broad, and the region spanned by
this DIB is plagued by stellar features. \cite{Snow4428} showed that
the intrinsic profile of the band is Lorentzian, and thus, rather than
numerically integrating the DIB profile, we preferred to fit a
Lorentzian profile to the observations and determine the equivalent
widths from the fitted parameters. In addition to our best fit, we
also determined Lorentzians that represent the upper and lower
envelope of the observed profiles; these then have a different
full-width-at-half-max (FWHM) and central depth (CD) value (while we
kept the same continuum). We found the difference between the best-fit
values and upper or lower envelope values and determined the
uncertainties on EW through error propagation.

\begin{figure}[h]
\begin{center}$
\begin{array}{crr}
           \includegraphics[trim={2cm, 6.5cm, 2cm, 5cm}, width=80mm]{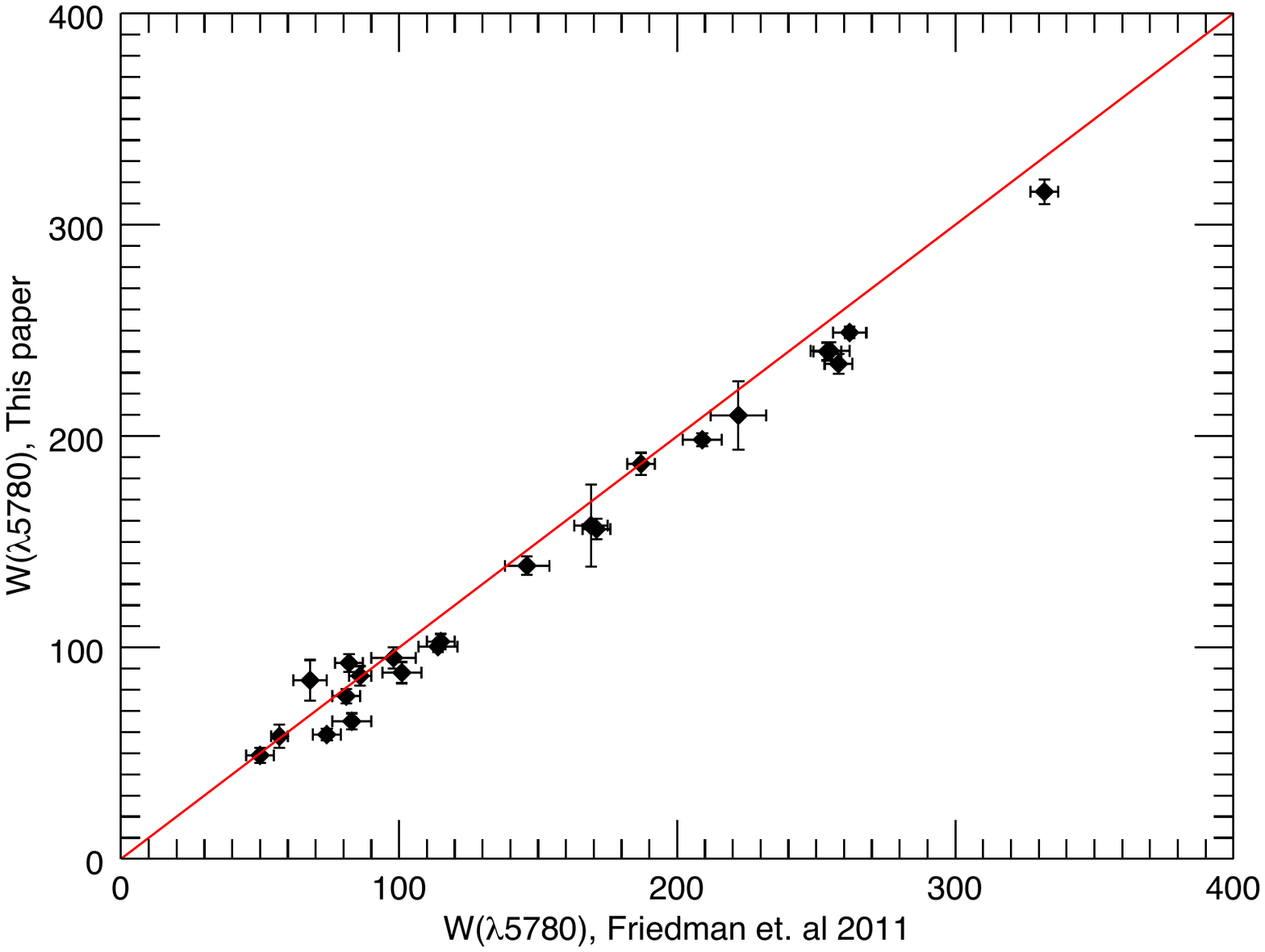}\\
           \includegraphics[trim={2cm, 6.5cm, 2cm, 5cm}, width=80mm]{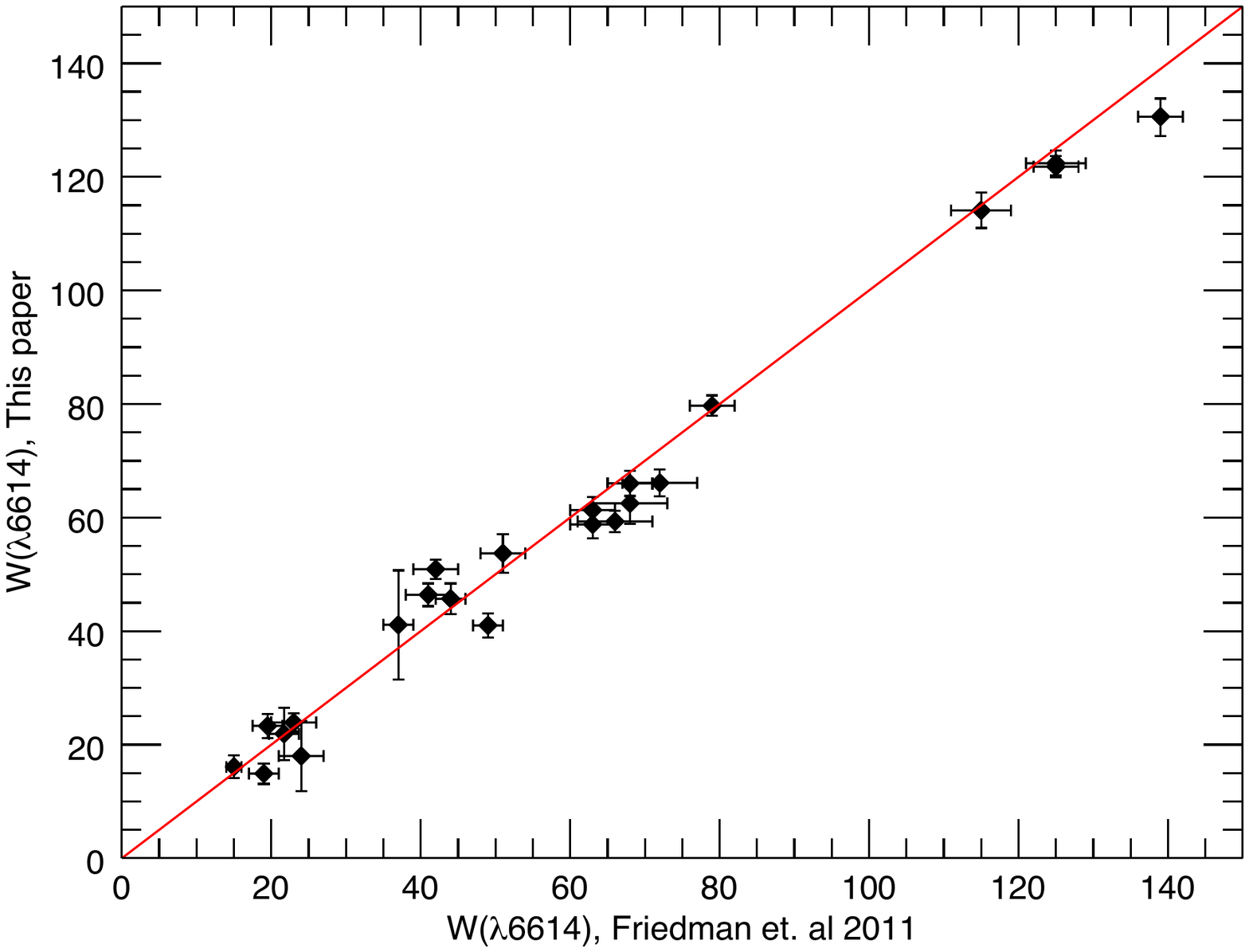}\\
\end{array}$    
\end{center}
\caption{\label{Fig:Friedman_comparison} A comparison of our
    measured $\lambda$5780 and $\lambda$6614 equivalent width values
    (vertical axes) to those found in \citet{Friedman2011} (horizontal
    axes). The two data sets agree very well; the correlation
    coefficient between our W($\lambda$5780) values is $r=0.995$, while
    that of W($\lambda$6614) is $r=0.993$.}
\end{figure}

We compared our measurements to values found in the literature
whenever possible and found a generally good agreement. For instance,
23 out of our 30 lines of sight were also studied by
\citet{Friedman2011} and we found a very good correlation between
their reported EW values and ours (see
e.g. Figure~\ref{Fig:Friedman_comparison}).

All of our equivalent width measurements are shown in
Table~\ref{table:EW_meas} in the appendix to this paper.

\floattable
\begin{deluxetable*}{lcccccccccccccc}
\tabletypesize{\small}
\rotate
\tablecaption{\label{table:target_info}Basic target data.}
\tablehead{
\colhead{Target} & \colhead{Alt.} & \colhead{RA} &
\colhead{DEC} & \colhead{V}  &
\colhead{E(B-V)} &
\colhead{N(\ion{H}{1})} &
\colhead{N(H$_2$)} &
\colhead{f(H$_2$)} &
\colhead{F$_\star$} &
\colhead{$\frac{{\rm W}(\lambda 5797)}{{\rm W}(\lambda 5780)}$} &
\colhead{$v_{\rm ISM}$\tablenotemark{a}} & \colhead{Ref\tablenotemark{a}}
& \colhead{Data}\\ 
\colhead{} & \colhead{Name} & \colhead{[J2000]}  & \colhead{[J2000]}
& \colhead{}  & \colhead{} & \colhead{[10$^{21}$cm$^{-2}$]}  &
\colhead{[10$^{20}$cm$^{-2}$]} & \colhead{}  &\colhead{} &\colhead{} &
\colhead{[km s$^{-1}$]} & \colhead{} & \colhead{Source}  
}
\startdata
\input{targettable}
\enddata

\tablecomments{RA and DEC are taken from SIMBAD. $V$, $E(B-V)$,
  N(\ion{H}{1}), N(H$_2$) and F$_\star$ values are from
  \citet{Jenkins} where we assume an uncertainty of $\pm$0.02 mag on
  the $E(B-V)$ values. $f(H_2)=2N(H_2)/[N(H) + 2N(H_2)]$ is the fraction of
  molecular hydrogen. }
\tablenotetext{a}{Value and reference refer to the velocity of the
  dominant interstellar component. } 
\tablerefs{(1) This paper; (2) \citet{WeltyHobbs2001}; (3) \citet{Bhatt2015}; (4) \citet{Welty1994}; (5) \citet{SWT2007}; (6) \citet{Adams1949}}
\end{deluxetable*}

%\clearpage
\subsection{Line of Sight Parameters}
\label{subsec:losparams}

To have a better chance of understanding what drives the variations in
the DIB strengths, we need to include other parameters that offer some
description of the lines of sight. The \citet{Jenkins} study
provides a critically reviewed set of measurements that describe the
lines of sight we study here and we reproduce some of their values in
Table~\ref{table:target_info}.

There are two quantities that are related to the dust in the line of
sight. The amount of dust is traditionally characterized by using the
color excess, E(B-V), and is often used as a normalization factor for
DIB strengths. We adopt an uncertainty of $\pm$0.02 mag for our E(B-V)
measurements. \citet{Jenkins} furthermore showed that the depletion of
different elements in a line of sight can be described by a single
parameter, $F_{\star}$, that is a measure for the ``total depletion'';
individual elemental depletion factors scale with $F_{\star}$. Since
depletion may play a role in the formation and/or destruction of the
DIB carriers, we include it here in our analysis.

Information on the gas in the line of sight stems from measurements of
hydrogen, in particular the column densities of neutral and molecular
hydrogen, N(\ion{H}{1}) and N(H$_2$), respectively. Note that total
hydrogen column densities N(H) are not listed, but can easily be
calculated as N(H) = N(\ion{H}{1}) + 2N(H$_2$). For HD~23630, a
reliable N(\ion{H}{1}) value could not be obtained by Jenkins;
consequently, he uses synthetically derived N(H) and F$_\star$ values,
which we adopt for this paper. We use this synthetic N(H) value and
the measured N(H$_2$) value to derive a synthetic N(\ion{H}{1}) from
N(H) $-$ 2N(H$_2$). For two other targets, HD~27778 and HD~202904,
Jenkins provides only upper limits and best values for
N(\ion{H}{1}). We take the lower limit to be zero in both cases. For
N(H), he lists only upper and lower limits. We adopted N(H)$_{\rm
  best}$ = N(\ion{H}{1})$_{\rm best}$ + 2 N(H$_2$)$_{\rm best}$ for
these targets, as well as synthetic F$_\star$ values. For HD~35149,
only an upper limit for N(H$_2$) is provided. In this case, N(H$_2$)
is quite small, so Jenkins takes N(H)~=~N(\ion{H}{1}) and calculates
F$_{\star}$ normally. Because N(H$_2$) is so small in this case, we
take Jenkins' value to be the best value, set the lower limit equal to
zero, and set the upper limit equal to the best value. It has been
suggested that f(H$_2$) can be used as an indicator for the amount of
interstellar UV radiation that can penetrate since H$_2$ is
dissociated when not shielded \citep[see e.g.][]{Cami1997,
  Sonnentrucker1997}. This could be an important parameter in our
study, and we therefore calculated the fraction of molecular hydrogen,
f(H$_2$), for each line of sight from N(H$_2$) and N(H). 

Similarly, it has long been known that the ratio
W($\lambda$5797)/W($\lambda$5780) is somehow related to the physical
conditions in a region of space \citep{Krelowski:MoreDibFamilies,
  Cami1997}; more specifically, it is also suggested to trace UV
exposure, where larger values of the W($\lambda$5797)/W($\lambda$5780)
ratio correspond to more sheltered ($\zeta$-type) environments. We
thus also include the W($\lambda$5797)/W($\lambda$5780) ratio as a
line of sight parameter in our study.

Our sample is quite diverse in terms of f(H$_2$): reported f(H$_2$)
values in diffuse clouds run from 0 to $\sim$0.8 \citep{Snow&McCall};
our sample covers a similar range from 0 to 0.824. Our sample also
covers a broad range of physical conditions, as it includes both
archetypal $\sigma$ and $\zeta$ sightlines. For comparison, our
W($\lambda$5797)/W($\lambda$5780) values range from 0.10 to 0.65 whereas
those calculated from EWs in a large sample of 133 stars by
\cite{Friedman2011} range from 0.15 to 0.85. Furthermore, the boundary
between $\sigma$ and $\zeta$ clouds is usually placed around
W($\lambda$5797)/W($\lambda$5780)=0.3 and therefore, we include a large
sample of both cloud types. For F$_\star$ we similarly cover much of
the possible range. Jenkins defines F$_\star$ such that a line of
sight with minimal depletions has a value of 0, and the
$-$15~km~s$^{-1}$ component of HD~149757 -- the archetype for {\em
  large} depletions -- has a value of 1.0. Our values for F$_\star$
are found between 0.37 and 1.19. For E(B-V), the situation is slightly
more complicated. Because we restrict ourselves to single clouds, we
are biased towards objects of lower reddening. Indeed, our E(B-V)
values range from 0.00 to 0.47, while it is not uncommon for
multiple-cloud sightlines to exceed 1. In particular, E(B-V) values
below 0.08 can be problematic; below this threshold, H$_2$ does not
exist in appreciable amounts \citep{Savage1977} and many DIBs
similarly fall below the limit of detection. Within our sample there are six
of these low-E(B-V) targets: HD~23630, HD~24760, HD~35715, HD~36822,
HD~143275, and HD~214680. Studies by \cite{lowEBV_1}, \cite{lowEBV_2},
and \cite{lowEBV_3} confirm that strong DIBs such as $\lambda$5780,
$\lambda$5797, $\lambda$6284, and $\lambda$6614 can be detected toward
such objects. Moreover, \cite{DIBs_low_EBV} confirm that
W($\lambda$5797)/W($\lambda$5780) is still sufficient to differentiate
between $\sigma$ and $\zeta$ environments, despite the weak
reddening. Some weaker DIBs may not appear in these poorly reddened
objects; in such cases, sightlines having larger E(B-V) values will
still be observable and will dominate the overall trends.

\newpage
\section{Principal Component Analysis}
\label{sec:PCA}

Several parameters are known to contribute to DIB strength; however,
most of these parameters are interrelated (e.g., a larger E(B-V)
implies larger column densities of all species). For this work, we
want to extract a set of uncorrelated parameters that best describe
observed DIB strengths. To achieve this goal, we perform a principal
component analysis (PCA).

PCA is a statistical technique used to analyze high-dimensional
data. The objectives of PCA are two-fold: (1) to reduce the number of
dimensions in a data set, and (2) to identify hidden patterns in a set
of data. We will primarily focus on the former in this analysis. For
derivations of PCA see \cite{PCA_Pearson} or \cite{PCA_Hotelling}; for
modern reviews, see \cite{PCA_Abdi} or \cite{PCA_Joliffe}; for uses of
PCA in astronomy see e.g. \citet{Yip:SDSS_PCA, Suzuki:quasarPCA,
  PCA_galaxies,2009MNRAS.394.1496B, PCA_quasarUV}.

In this Section, we first define some of the terminology and notation
associated with PCA and provide a basic mathematical overview of the
process. We then discuss some of the problems and underlying
assumptions associated with PCA, and how we address each of
them. Finally, we provide a simple example using two variables to
better illustrate our methodology and facilitate the interpretation of
our results.

\subsection{Definitions and Terminology}
\label{subsec:PCA_definitions}

The starting point of our analysis is a set of $n$ variables for which
we have $m$ measurements $x_i$; in our case, we use $n=23$ variables
(representing DIB equivalent widths and line of sight parameters) that
are measured for $m=30$ lines of sight. Our measurements thus span an
$n$-dimensional parameter space -- each dimension corresponding to a
different variable -- and a full set of measurements for a single line
of sight can be represented by an $n\times1$ vector in this parameter
space.  Starting from this set of \textit{n} possibly correlated (and
thus not necessarily orthogonal) variables, a PCA finds a coordinate
transformation that casts the variables in terms of a new, {\em
  orthogonal}, $n$-dimensional reference frame:
\begin{equation}
\label{Eq:PC_vector}
y_i~\hat{n}^{\prime}_i = a_{i,1} x_1 \hat{n}_1 + a_{i,2}
x_2 \hat{n}_2 + ... + a_{i,n} x_n \hat{n}_n
\end{equation}
where the $\hat{n_i}$ represent unit vectors in the original parameter
space and $\hat{n}^{\prime}_i$ unit vectors in the new reference frame
-- these are the so-called principal components (PCs).  The
coefficients, $a_{i,j}$, are constrained such that their squared sums
equal one:

\begin{equation}
a_{i,1}^2 + a_{i,2}^2 + ... + a_{i,n}^2 = 1
\end{equation}

The coordinate transformation is chosen such that when representing
the measurements in the new reference frame, the largest amount of
variance in the data occurs along the axis defined by the first PC,
and the second PC accounts for as much of the remaining variation in
the data as possible, with the constraint that it is orthogonal to
PC$_1$. Each successive PC has the same properties: they are linear
combinations of the original variables, and each accounts for as much
of the remaining variation as possible, while being orthogonal to (and
thus, uncorrelated with) all previous PCs.

The entire coordinate transformation can be written in matrix
notation:
\begin{equation}
\label{eqn:PCA_matrix_eqn}
Y=AX
\end{equation}
where $X$ is an $n\times m$ matrix containing the original set of
data; $A$ is the $n\times n$ transformation matrix; and $Y$ is an
$n\times m$ matrix containing the transformed data points in the new
reference frame defined by the set of PCs\footnote{Note that in the
  PCA analysis by \citet{Suzuki:quasarPCA}, their Eq.~1 corresponds to
  the inverse of our Eq.~\ref{eqn:PCA_matrix_eqn}.}.

The PCA obtains the transformation matrix $A$ from the eigenvalues
($\lambda_1$, $\lambda_2$, ..., $\lambda_n$) of the covariance matrix
of the original variables. The rows of this matrix $A$ are the
corresponding eigenvectors; each eigenvector describing the
projections of the original variables onto the new coordinate system
defined by the PCs.  The eigenvalues are furthermore ordered such that
$\lambda_1$ is the largest and $\lambda_n$ is the smallest with
$\sum_i^n \lambda_i = n$. Each eigenvalue indicates how many original
variables a single PC can account for, i.e., an eigenvalue of 2
indicates that the corresponding PC carries the same weight as two of
the original variables.

One of they key features of a PCA is the ability to reduce the number
of dimensions in a data set. Indeed, it is often the case that we can
use the results of a PCA to accurately describe the variation in a
data set with a reduced number of variables. As such, PCs that account
for very little variation are often ignored, and only the leading
\textit{p} components are kept. Selecting a value for \textit{p} is
somewhat arbitrary, but the decision is usually based on one of three
criteria: (1) once a certain amount of variation is accounted for
(e.g., 90\%), all further PCs are ignored; (2) only PCs with
eigenvalues greater than one are kept (representing variables that can
replace more than one of the original variables); or, (3) if an
``elbow" (a sharp, sudden drop-off) is noticed in a so-called
``screeplot" (i.e., a plot of eigenvalue vs. component number, see
e.g. Fig.~\ref{fig:screeplot}), all PCs beyond the elbow are ignored
\citep{PCA_Joliffe}. If \textit{p} components are kept, then the last
$n - p$ rows of A are eliminated, and Y becomes a $p\times m$
matrix. At that point, we effectively express the original $n$
variables with only $p$ new variables, while keeping as much of the
variance in the data as desired.

\subsection{Treatment of Data}
\label{subsec:PCA_data_treatment}

The variables $x_i$ that we want to include in our analysis are listed
in column 1 of Table~\ref{table:Scaling_factors}. Since PCA is
concerned with linear combinations of variables, it is redundant to
include N(H) in addition to N(\ion{H}{1}) and N(H$_2$); however,
whereas N(\ion{H}{1}) and N(H$_2$) measure the column densities of
individual species, N(H) approximates the total amount of gas in the
line of sight. Recognizing this distinction, we choose to keep N(H) in
addition to N(\ion{H}{1}) and N(H$_2$).

There are a few underlying assumptions associated with PCA, which we
will discuss. The first is that raw data is comparable in units and
magnitude; otherwise, the results of PCA will be more strongly
influenced by variables that have larger variances (i.e., variations
among N(H) values which are on the order of 10$^{21}$ will completely
dominate those of E(B-V), which are on the order of 10$^{-1}$). To
address this problem, we standardize each variable prior to performing
PCA:
\begin{equation}
\label{eqn:standardization}
z_{i,j} = \frac{x_{i,j} - \bar x_i}{s_{x_i}}
\end{equation}
where $i$ refers to the $i^{th}$ variable and $j$ refers to the
$j^{th}$ observation. Essentially, we subtract the mean and divide the
residuals by the standard deviation, and use the resulting variables
$z_{i,j}$ as input to the actual PCA. Columns 2 and 3 of
Table~\ref{table:Scaling_factors} show the mean and standard deviation
of each of our variables, respectively. After this process, the
standardized variables ($z_1$, $z_2$, ... $z_{23}$) each have a mean
of zero and a standard deviation of one.

A second assumption of PCA is that all values are accurately
known. This assumption is challenged by the fact that our measurements
have uncertainties. Since PCA does not consider uncertainties, we need
a way to quantify how reliable our results are. Since the PCs
represent vectors in a multi-dimensional parameter space, this is not
straightforward.

We therefore used a Monte Carlo (MC) simulation to estimate the
uncertainties on our results. We first performed PCA using our
measured values. These results provide the PCs that we discuss in the
sections that follow. Next, we generated 1,000 perturbed data sets, by
adding random noise to each observation, selected from a normal
distribution with a mean of zero and a standard deviation
corresponding to the error bar on each measurement. By definition,
none of our measured quantities can be negative, so if a perturbed
measurement was negative after the addition of random noise, the value
was replaced with zero. For each of these new data sets, we
standardized the variables \footnote{Note that we recalculated the
  means and standard deviations for each perturbed data set before
  standardizing the variables.} and performed a separate PCA. The
result is 1,000 unique transformation matrices. For each entry in the
transformation matrix, we found the mean ($\bar a_{i,j}$) and standard
deviation ($s_{a_{i,j}}$) among the 1,000 unique matrices. Then we
computed the upper and lower limit on each value in the matrix A
according to $\bar a_{i,j} \pm s_{a_{i,j}}$. We compared these upper
and lower limits to the original, unperturbed results and ultimately,
were left with positive and negative errors on each entry in the
transformation matrix (i.e., each component of each eigenvector). We
applied the same methods to obtain uncertainty measurements on each
entry in Y, the matrix of transformed data points.

\begin{deluxetable}{lDD}
\tablecaption{\label{table:Scaling_factors}Input variables used in the
  PCA, and their mean values and standard deviations. } 
\tabletypesize{\small}
\tablehead{
\colhead{Variable} & \twocolhead{Mean}  & \twocolhead{Standard}\\
\colhead{Name} & \twocolhead{}  & \twocolhead{Deviation} \\
\colhead{($x_i$)} & \twocolhead{($\bar x_i$)}  &
\twocolhead{($s_{x_i}$)}
}
\decimals
\startdata
$x_1 = {\rm W}(\lambda 4428)$ & 645.9 & 350.9 \\
$x_2 = {\rm W}(\lambda 4964)$ & 6.8 & 5.7 \\
$x_3 = {\rm W}(\lambda 5494)$ & 5.6 & 4.1 \\
$x_4 = {\rm W}(\lambda 5513)$ & 3.9 & 3.9  \\
$x_5 = {\rm W}(\lambda 5545)$ & 5.9 & 4.3 \\
$x_6 = {\rm W}(\lambda 5546)$ & 2.8 & 2.2 \\
$x_7 = {\rm W}(\lambda 5769)$ & 2.6 & 2.2 \\
$x_8 = {\rm W}(\lambda 5780)$ & 131.5 & 77.0 \\
$x_9 = {\rm W}(\lambda 5797)$ & 37.7 & 27.8 \\
$x_{10} = {\rm W}(\lambda 5850)$ & 15.6 & 13.6 \\
$x_{11} = {\rm W}(\lambda 6196)$ & 13.5 & 8.3 \\
$x_{12} = {\rm W}(\lambda 6270)$ & 20.7 & 13.9 \\
$x_{13} = {\rm W}(\lambda 6284)$ & 149.4 & 84.1 \\
$x_{14} = {\rm W}(\lambda 6376)$ & 9.3 & 7.4 \\
$x_{15} = {\rm W}(\lambda 6379)$ & 24.7 & 18.0 \\
$x_{16} = {\rm W}(\lambda 6614)$ & 52.5 & 35.2 \\
$x_{17}$ = E(B-V) &               0.20 & 0.13 \\
$x_{18}$ = N(\ion{H}{1}) & 1.0$\times 10^{21}$ & 9.2$\times 10^{20}$ \\
$x_{19}$ = N(H$_2$)      & 2.4$\times 10^{21}$ & 2.6$\times 10^{20}$ \\
$x_{20}$ = N(H)          & 1.5$\times 10^{21}$ & 1.2$\times 10^{21}$ \\
$x_{21}$ = f(H$_2$)      & 0.27 & 0.23 \\
$x_{22}$ = F$_{\star}$     & 0.75 & 0.21 \\
$x_{23}$ = $\frac{\textrm{W}(\lambda5797)}{\textrm{W}(\lambda5780)}$ &
0.2907 & 0.1567 \\
\enddata	
\tablecomments{Entries for $i$=1,2,...16 refer to equivalent widths of
  named features in m\AA. Units for the remaining entries are those
  specified in Table~\ref{table:target_info}. Standardized variables
  $z_i$ correspond to the $x_i$, e.g. $z_8$ is the standardized
  equivalent width of the $\lambda5780$ DIB. }
\end{deluxetable}

A third assumption associated with PCA is that each variable follows a
normal distribution. We tested our variables for normality and found
that in some cases, the distributions were not normal. The assumption
of normality, however, is not critical and the results of PCA should
still be valid if this condition is not met, so we decided to continue
with the analysis.

Finally, the results of PCA can be sensitive to outliers. Despite
including some unusual targets (e.g., HD~147933 is found within a dark
cloud, and HD~36822 and HD~36861 are found near \ion{H}{2} regions),
we do not remove any targets from our sample. In order to probe a wide
range of interstellar conditions, it is important to include these
abnormal lines of sight: Any conclusions made about the DIBs through
this analysis should extend to these sightlines as well. The only
potential complication is for the six Be stars discussed in
Section~\ref{subsec:targets}. The atypical E(B-V) values arise from
circumstellar material and are therefore not a part of the
interstellar DIB environment. We acknowledge that they could influence
our results and therefore we differentiate these points in the plots
and analyses that follow.

\subsection{A Simple 2D Example}
\label{subsec:2D_example}

It is insightful to illustrate our methodology by performing a PCA on
a simple, two-dimensional dataset; we therefore performed a PCA
analysis using only the measurements for the variables E(B-V) and
N(H). Using the notation provided in
Table~\ref{table:Scaling_factors}, we denote these variables $x_{17}$
and $x_{20}$ and their corresponding standardized forms as $z_{17}$
and $z_{20}$. The $2 \times 30$ matrix of (standardized) measurements
X from Equation~\ref{eqn:PCA_matrix_eqn} is then:

% letter c is showing up in matrix for some reason
\begin{equation}
X = \begin{bmatrix}
z_{17} \\
z_{20}
\end{bmatrix}\\
=\begin{bmatrix}
0.273 & -0.106 & \dots & -0.333 \\
0.121 & 0.121 & \dots & -0.285
\end{bmatrix}
\end{equation}

From this, the PCA results in the following $2 \times 2$
transformation matrix A:

\begin{equation}
A = \begin{bmatrix}
a_{1,1} & a_{1,2}\\
a_{2,1} & a_{2,2}
\end{bmatrix}
 = \begin{bmatrix}
0.707 & 0.707\\
0.707 & -0.707
\end{bmatrix}
\end{equation}

as well as the $2 \times 30$ matrix, Y, containing the transformed
data points:

\begin{equation}
Y = \begin{bmatrix}
PC_{1} \\
 PC_{2}
\end{bmatrix}
= \begin{bmatrix}
0.279 & 0.011 & \dots & -0.437\\
0.107 & -0.161 & \dots & -0.034 \\
\end{bmatrix}
\end{equation}

Note that we use the notation $PC_i$ to describe the measurement
values in the new coordinate frame described by the PCs. That is,
$PC_i = y_i$ in Eq.~(\ref{Eq:PC_vector}).  The results pertaining to
the PCs for this example are summarized in table
\ref{table:2D_results}. Column 1 lists the PC number, while Column 2
shows the eigenvalue associated with that PC. Since there are two
original variables, the eigenvalues sum to 2. The fraction of the
total variation in the data for which each PC is responsible is shown
in Column 3, followed by the cumulative fraction in Column 4. The
(unit length) eigenvectors are shown in Column 5.

\begin{deluxetable}{crrrrr}
\tablewidth{0pt}
\tablecaption{\label{table:2D_results} Principal components,
  eigenvalues and relative importance of each PC for a 2D example involving E(B-V) and N(H).} 
\tablehead{
\colhead{PC} & \colhead{Eigen-} & \colhead{\%}  & \colhead{Cumulative} & \colhead{Eigenvector}  \\
\colhead{} & \colhead{value} & \colhead{Variation}  & \colhead{\%} & \colhead{} }
\startdata
	1  &1.813 &90.63 & 90.63 & (0.707, 0.707)\\
	2  &0.187 &9.37 & 100.00 & (0.707, -0.707)\\
\enddata
\end{deluxetable}

\begin{figure}[t]
\begin{center}$
\begin{array}{crr}
           \includegraphics[width=80mm,trim={1.5cm, 6.1cm, 1.5cm, 6.1cm}]{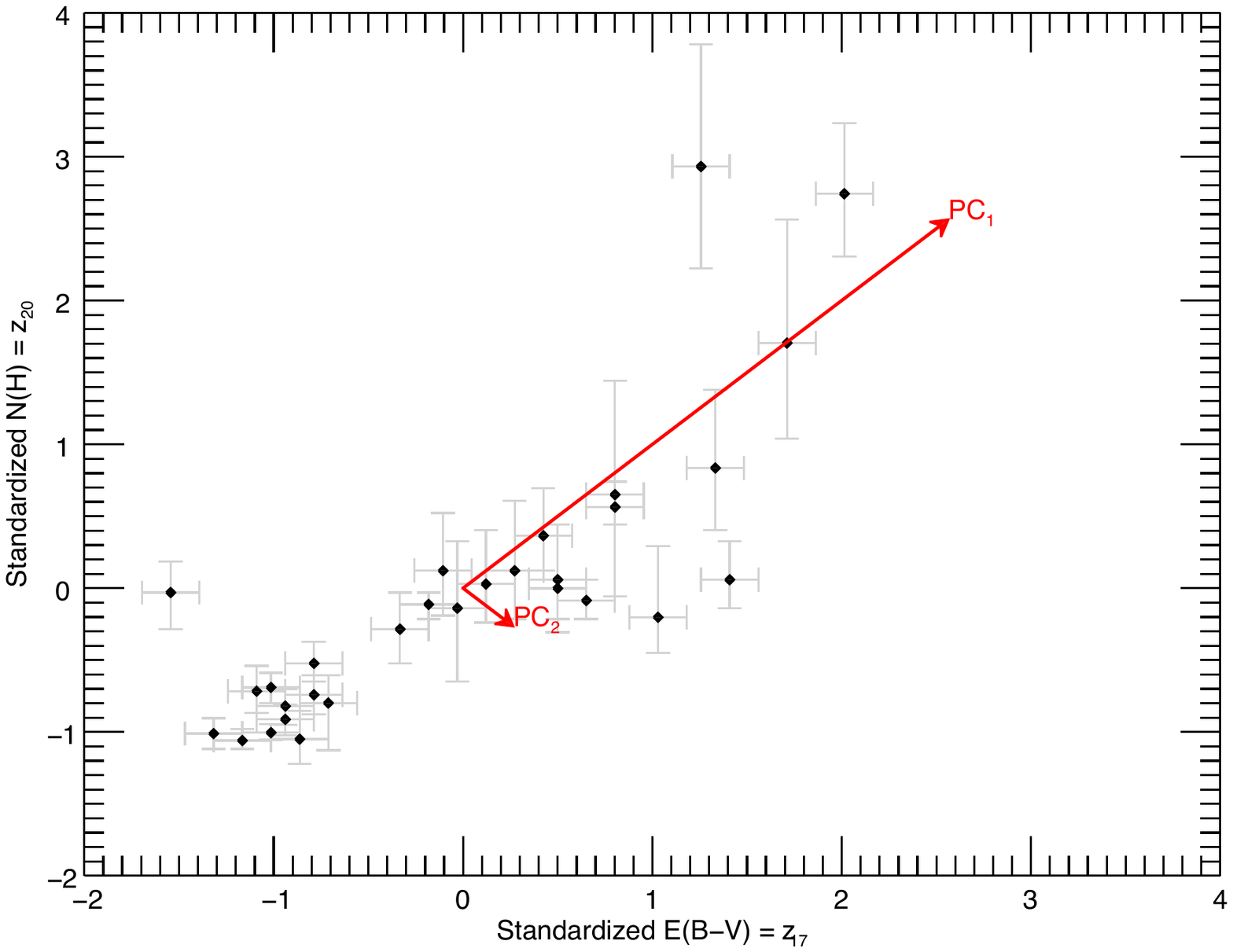}\\
           \includegraphics[width=80mm, trim={1.5cm, 6.1cm, 1.5cm, 6.1cm}]{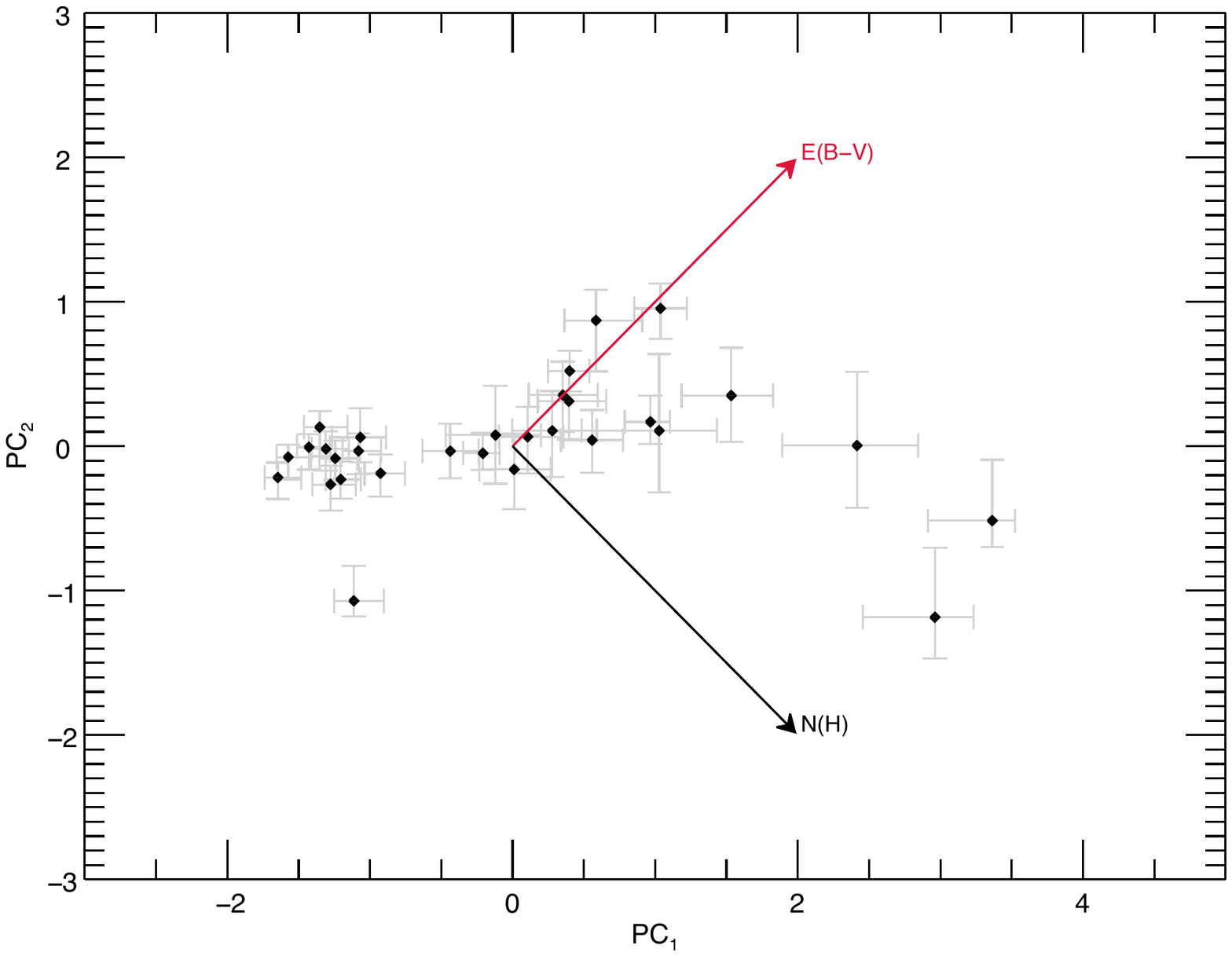}\\
\end{array}$    
\end{center}
\caption{\label{fig:2D_biplot} (\textit{Top}:) The original
    standardized data points and their uncertainties (black) and the
    PCs (red), the latter scaled up by a constant factor for
    clarity. PC$_1$ points in the direction of maximum
    variation. PC$_2$ is perpendicular to PC$_1$ and accounts for the
    remaining variation.  \\ (\textit{Bottom}:) A so-called ``biplot''
    illustrating the relationship between the original variables and
    the new set of PCs. The figure shows the transformed data points
    (Y) in the PC$_1$-PC$_2$ plane. The original variables are shown
    as vectors projected onto the PC$_1$-PC$_2$ plane. For example,
    E(B-V) has a projection on PC$_1$ equal to $a_{1,1}=0.707$ and
    projectio\ on PC$_2$ equal to $a_{1,2}=0.707$. Similarly, N(H) has
    a projection on PC$_1$ equal to $a_{2,1}=0.707$ and a projection
    on PC$_2$ equal to $a_{2,2}=-0.707$. Note that the vectors have
    been scaled for clarity in this figure too. }  
\end{figure}

Figure \ref{fig:2D_biplot} illustrates these results in two ways. In
the top panel, we show the original data set along with the PC vectors
(i.e., the eigenvectors corresponding to PC$_1$ and PC$_2$). From this
figure, it is clear that the vector PC$_1$ lies in the direction of
maximum variance. From table \ref{table:2D_results}, we see that in
fact almost all of the variation in the data set ($>$90\%) is
accounted for by PC$_1$. This indicates that a single variable can
adequately describe variations among E(B-V) and N(H) -- an
interstellar finding that is well known \citep[e.g.][see also
  discussion below]{BSD1978}.

In the bottom panel of Figure \ref{fig:2D_biplot}, we illustrate the
results in the form of a so-called ``biplot". Here, we show the
transformed data points (i.e., those of matrix Y) and the projections
of the original variables onto the PC$_1$-PC$_2$ plane. This format is
much more convenient for interpreting the PCs. For instance, we can
examine the projections of E(B-V) and N(H) onto PC$_1$ (i.e., only the
horizontal components of the vectors) to see how PC$_1$ influences
each original variable. Both projections onto PC$_1$ are positive;
thus, an increase of PC$_1$ implies an increase of both E(B-V) and
N(H). An obvious interpretation for PC$_1$ is therefore the amount of
material in the line of sight: more material will result in larger
E(B-V) and N(H) values at the same time. 

Doing the same for PC$_2$, we see that the vertical component of the
E(B-V) vector is positive, while that of N(H) is negative. PC$_2$
therefore represents some difference between E(B-V) and N(H). A
possible interpretation is therefore the dust-to-gas ratio: an
increase in the dust-to-gas ratio can result in an increase in the
E(B-V) or a decrease in the N(H).

Using the eigenvectors, we can construct equations for PC$_1$ and
PC$_2$. Since the PCs were constructed from the standardized
variables, the resulting equations will be in terms of $z_{17}$ and
z$_{20}$; using Eq.~\ref{eqn:standardization} and the mean and
standard deviations in Table~\ref{table:Scaling_factors}, however, we
can express them in terms of the original variables, E(B-V) and N(H).

\begin{align}
\label{eqn:2D_PC1}
PC_1=&0.707(z_{17})+0.707(z_{20}) \\
=&  5.35(E(B-V)) + (5.93\times 10^{-22})N(H) - 1.99 \nonumber
\end{align}

\begin{align}
\label{eqn:2D_PC2}
PC_2=&0.707(z_{17})-0.707(z_{20}) \\
=& 5.35(E(B-V)) - (5.93\times 10^{-22}) N(H) - 0.19 \nonumber
\end{align}

We can also use our example to demonstrate how to reconstruct the
observed data with reduced dimensionality. In our 2D example, PC$_1$
accounts for most of the variation in the data, and one could argue
that the variations in PC$_2$ are mostly insignificant. In other
words, we can attribute variations in PC$_2$ to noise in the data, and
thus we can describe our data by just using 1 variable:
PC$_1$. Mathematically, we thus eliminate the final column of matrix
A, forming a $1\times30$ matrix, Y; in doing so, we leave
Equation~\ref{eqn:2D_PC1} as is and set Equation~\ref{eqn:2D_PC2}
equal to zero. Solving for N(H) then reveals: 

\begin{align}
\label{eqn:Solve_NH_2D}
%N(H) = & (9.03 \times E(B-V) - 0.33) \times10^{21}\\
N(H) = & 9.0 \times 10^{21} E(B-V) - 0.3 \times10^{21}
\end{align}

while a least-squares fit to the original data set yields

\begin{equation}
\label{eqn:NH_EBV_bestfit}
%N(H) = (9.96 E(B-V) - 0.50) \times 10^{21}
N(H) = 10.0 \times 10^{21} E(B-V) - 0.50 \times 10^{21}
\end{equation}

\begin{figure}[t!]
\begin{center}
\includegraphics[width=8.5cm,  trim={1.5cm, 6.1cm, 1.5cm, 6.1cm}]{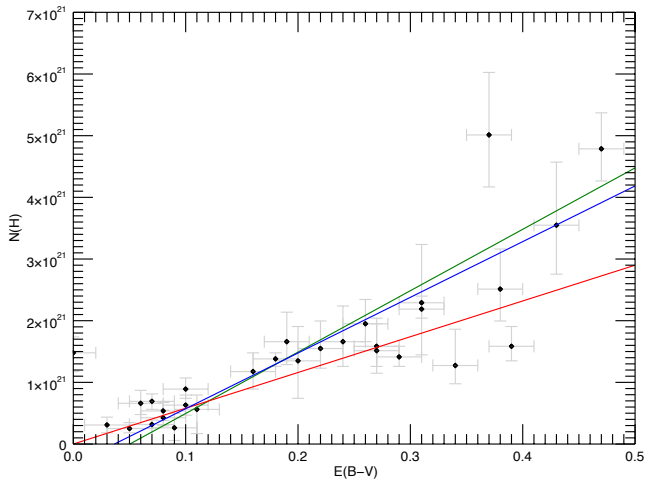}
\caption{\label{fig:NH_EBV_Comparison} N(H) (in cm$^{-2}$) as a
  function of E(B-V). Equation~\ref{eqn:Solve_NH_2D}, obtained by
  setting PC$_2$=0 is represented by the blue line. The best-fit line
  obtained through a linear least squares fit
  (Equation~\ref{eqn:NH_EBV_bestfit}) is shown in green. The
  well-known relation found by \citet[][Eq.~\ref{eqn:NH_EBV}]{BSD1978}
  is shown in red for comparison.}
\end{center}
\end{figure}

Both results are comparable (see the blue and green line in
Fig.~\ref{fig:NH_EBV_Comparison}), illustrating that a PCA can be used
to reliably reconstruct relationships between variables that are
present in the original data set. Note though that this is somewhat
different from the well-known finding by \cite{BSD1978}:

\begin{equation}
\label{eqn:NH_EBV}
N(H) = 5.8 \times 10^{21} E(B-V)
\end{equation}

This relation is also shown in Fig.~\ref{fig:NH_EBV_Comparison} (red
line). Two important factors contribute to the difference between
\citet{BSD1978} and our results. First, our sample of sightlines is
biased to contain only single cloud lines of sight, with diverse
physical conditions whereas \cite{BSD1978} used a large sample of much
more reddened lines. Second, the \citet{BSD1978} relation was forced
to go through the origin; the PCA and least squares fit were not. A
discussion of such differences is not relevant for our work though.

% \begin{deluxetable}{crrr}
% \tablewidth{0pt}
% \tablecaption{\label{table:NH_EBV_slope_int} Comparison of slope and intercept for N(H) as a function of E(B-V)} 
% \tablehead{
% \colhead{Fit Method} & \colhead{Slope ($\pm 1\sigma$)} & \colhead{Intercept ($\pm 1 \sigma$)} }
% \startdata
% 	Linear least squares fit & ($9.96 \pm 0.39) \times 10^{21}$ & $(- 5.0 \pm 0.88) \times 10^{20}$ \\
% 	Setting PC$_2 =  0$ & $9.03 \times 19^{21}$ & $ -3.3 \times 10^{20}$
% \enddata
% \end{deluxetable}

This example shows the steps involved in carrying out a PCA and using
its results. For what follows, we will be using multi-dimensional data
though; the only complication is then that for the discussion and
interpretation, we need to work with projections onto a 2D space for
proper visualization.

\section{PCA Results}
\label{sec:full_PCA}

In this section, we perform PCA using the full set of 23 variables
listed in Table~\ref{table:Scaling_factors}. The resulting PCs are
listed in Table~\ref{table:PCA_results}, along with their
corresponding eigenvalues and the percent of the variation in the full
data set for which each PC is responsible. Decidedly fewer than 23
components are required to describe the observed variations; in fact,
PC$_{23}$ is completely unnecessary, indicating that there is a
perfect correlation between some of our variables (i.e., that between
N(H), N(\ion{H}{1}), and N(H$_2$)). Most of the variation is captured
by the first few PCs; PC$_1$ alone accounts for over 65\%, while
PC$_1$ and PC$_2$ collectively account for over 80\% of the overall
variation.

\begin{figure}[t!]
\begin{center}
\includegraphics[width=8.5cm,  trim={2cm, 9.5cm, 2cm, 9cm}]{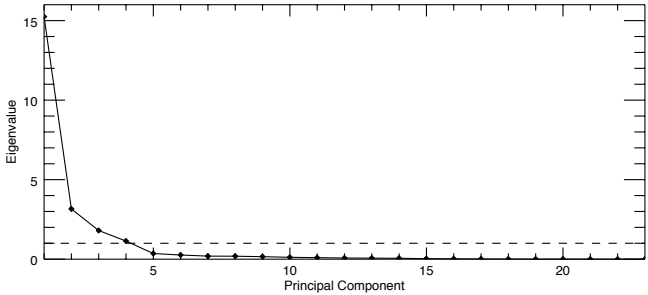}
\caption{\label{fig:screeplot} A screeplot illustrating the relative
  importance of each PC. The dashed line indicates an eigenvalue of
  one. The first four PCs lie above this limit.} 
\end{center}
\end{figure}

To determine \textit{p}, the number of significant PCs, we use
criteria 2 and 3 from Section~\ref{subsec:PCA_definitions}. A
screeplot is shown in Figure~\ref{fig:screeplot}, where the dashed
line illustrates an eigenvalue of one. Only the first four PCs lie
above this limit. For the subsequent PCs, the uncertainties obtained
through our MC simulation become too large to be interpreted. This
could indicate that there are no clear relations beyond this point
(i.e., the variation is guided by MC perturbations rather than
systematic trends in the data) or could be the result of compounding
uncertainties (i.e., since PCs are constrained to be orthogonal, each
PC has larger uncertainties than all previous PCs).

\begin{deluxetable}{crrrr}
\tablewidth{0pt}
\tablecaption{\label{table:PCA_results} Principal components,
  eigenvalues and relative importance of each PC. } 
\tablehead{
\colhead{PC} & \colhead{Eigenvalue} & \colhead{\% Variation}  & \colhead{Cumulative \%}   }
\startdata
	1  &15.248 &66.30 & 66.30 \\
	2  & 3.158 &13.73 & 80.03 \\
	3  & 1.801&7.83& 87.86 \\
	4  & 1.139 &4.95 & 92.81 \\
	5  & 0.355 &1.54 & 94.35 \\
	6  & 0.262 &1.14 & 95.49 \\
	7  & 0.192 &0.84 & 96.33 \\
	8  & 0.186 &0.81 & 97.14 \\
	9  & 0.157 &0.68 & 97.82\\
	10 & 0.117 &0.51 & 98.33\\
	11 & 0.096 &0.42 & 98.75\\
	12 & 0.074 &0.32 & 99.07\\
	13 & 0.066 &0.29 &  99.35\\
	14 & 0.055 &0.24 & 99.60\\
	15 & 0.032 &0.14 & 99.74 \\
	16 & 0.025 &0.11 & 99.85 \\
	17 & 0.012 &0.05 & 99.90 \\
	18 & 0.008 &0.03 & 99.93 \\
	19 & 0.006 &0.03 & 99.96 \\
	20 & 0.005 &0.02 &99.98 \\
	21 & 0.003 &0.01 &99.99\\
	22 & 0.002 &0.01 &100.00\\
	23 & 0.000 &0.00 &100.00\\
\enddata
\end{deluxetable}

Rather than listing the individual components of each PC, we present
our results in the form of biplots (e.g., see
Figure~\ref{fig:PC1-PC2}), although we include the full equations for
the first four PCs in the Appendix (see Equations \ref{eqn:PC1_eqn}
through \ref{eqn:PC4_eqn}). We use our MC simulation described in
Section~\ref{subsec:PCA_data_treatment} to draw error bars on these
data. The original set of axes (i.e., the original variables) are also
illustrated as vectors projected onto the two-dimensional PC
plane. These are obtained through the transformation matrix, A (i.e.,
column $i$ of matrix A contains the projections of the $n$ original
variables onto PC$_i$). Our MC approach also provides uncertainties on
these values, which we illustrate as rectangular outlines surrounding
each vector head.

Below, we take a close look at the results of our PCA, and try to
interpret the PCs in terms of the physical parameters that drive the
variations in the DIBs. However, while each PC represents a single
measurable parameter that drives part of the variations, this
parameter can be related to more than one \textit{physical} quantity;
if so, these physical quantities must then necessarily be
correlated. There is thus some risk in trying to interpreting the PCs
in terms of single quantities. To facilitate our interpretation, we
have furthermore performed two additional PCAs -- one using only DIBs
($z_1$, $z_2$, ..., $z_{16}$) and using only line of sight parameters
($z_{17}$, $z_{18}$, ..., $z_{23}$); we compare the results to those
obtained by including the full set of 23 variables. Each PC will be
discussed separately in the sections that follow.

\begin{figure*}
\centering
\resizebox{\textwidth}{!}{\includegraphics[trim=1.5cm 3cm 1.5cm 3.5cm]{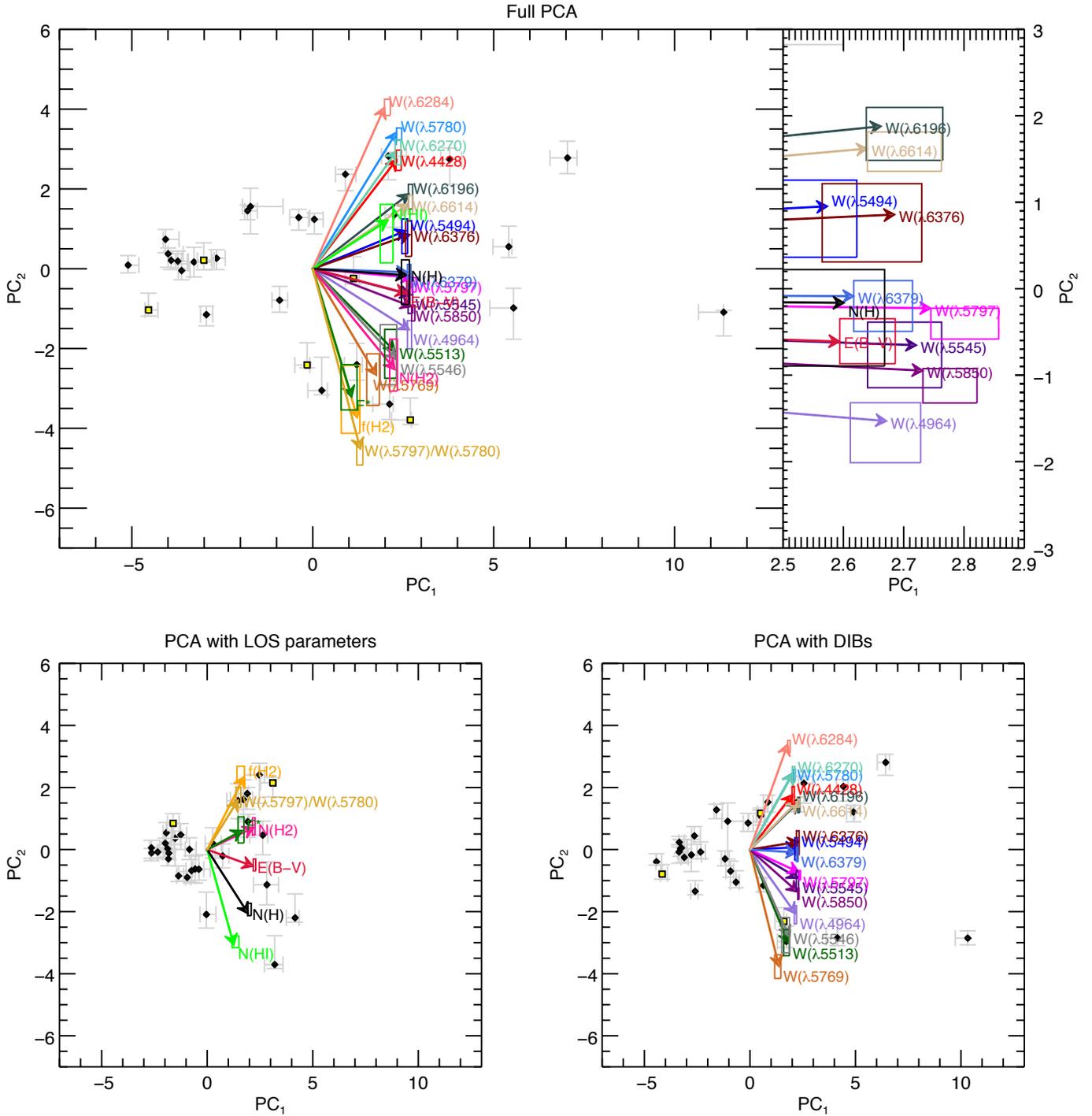}}
\caption{\label{fig:PC1-PC2}PC$_1$-PC$_2$ biplots for (\textit{Top}:)
  all variables, (\textit{Bottom left}:) line of sight parameters only
  (excluding DIBs), and (\textit{Bottom right}:) DIBs only, excluding
  line of sight parameters. The 23-dimensional vectors corresponding
  to the original variables are projected onto the PC$_1$-PC$_2$
  plane. The rectangular outlines surrounding the vectors indicate the
  uncertainty range for each projection, obtained through a MC
  simulation. Note that the vectors have been scaled by a constant
  factor for better visualization (as was done in
  Fig~\ref{fig:2D_biplot}).  Be stars are indicated by yellow
  squares. To help with clarity, a zoomed-in and rescaled portion of
  the full PCA results is presented next to the main figure. The same
  color scheme is used for all figures.}
\end{figure*}

\subsection{PC$_1$: DIB column density}

Biplots for PC$_1$ and PC$_2$ are shown in Figure~\ref{fig:PC1-PC2};
the top panel shows the full PCA results while the other panels show
the results of the additional PCA's using only line of sight
parameters and using only DIBs. As was the case for the 2D example
(see Sect.~\ref{subsec:2D_example}), the biplot shows the measurements
for all lines of sight transformed into the PC reference frame, and
only shows the values for the first two PCs -- i.e. the two parameters
that together carry 80\% of the variance in the data set. The biplot
also shows the projection of the original variables onto the plane
defined by these two PCs. Such projections are useful to find trends
and correlations: the larger the projection of an original variable is
onto an axis, the better the original variable correlates with the
PC. Furthermore, original variables whose projections onto the PC
plane are similar, must necessarily represent tight correlations as
well. Note as well the rectangular areas at the vector heads (best
seen in the close-up in the top panel to the right) that represent the
uncertainties on the projections as derived from a MC simulation.

Since PC$_1$ and PC$_2$ capture most of the variation in the data, the
PC1-PC2 biplot is particularly insightful for identifying trends and
correlations. For example, W($\lambda$6614) and W($\lambda$6196) align
almost perfectly with one another within their respective uncertainty
ranges in PC$_1$. This implies that the two DIBs must correlate very
tightly -- a well-known fact for these DIBs \citep[see
  e.g.][]{McCall2010}. Similarly, W($\lambda$5780) is known to
correlate well with W($\lambda$6284), W($\lambda$6270), and
W($\lambda$4428) \citep{Herbig1993, 3families} and these appear
grouped together on the PC$_1$-PC$_2$ plane. Many more known relations
between DIB strengths \citep[see e.g.][for an overview]{Lan:SDSS_DIBs}
can be recovered from these plots.

To infer clues about what PC$_1$ physically represents, we can look at
the projections of the original variables onto the PC$_1$ axis. These
projections can be easily determined from Eqs.~(\ref{eqn:PC1_eqn}) and
(\ref{eqn:PC2_eqn}). For example, the vector that corresponds to the
projection of the $\lambda$5780 DIB (i.e. $z_8$, see
Table~\ref{table:Scaling_factors}) onto the PC1-PC2 plane is given by
the coefficients in front of $z_8$ in Eqs.~(\ref{eqn:PC1_eqn}) and
(\ref{eqn:PC2_eqn}), i.e. the corresponding vector is (0.207,
0.306). This vector is shown (albeit scaled up for visualization) in
Fig.~\ref{fig:PC1-PC2}.  As PC$_1$ increases, so too does each of the
original variables. Therefore, we interpret PC$_1$ as measuring the
amount of material in the line of sight; as the amount of material
increases, so does the strength of each DIB along with variables such
as color excess and the various column densities. Intuitively, this
makes sense: The factor that most strongly contributes to the column
density of any species in a given line of sight is the amount of
material.

We can elaborate further by examining the other two PCA cases. The
bottom right panel of Figure~\ref{fig:PC1-PC2} shows the PCA results
when only DIBs are considered. Again, each variable increases in the
same direction, so PC$_1$ must once again be some proxy for the amount
of material in the line of sight. However, since we only consider DIBs
in this case, PC$_1$ is specifically tracing the amount of material
that contributes to DIBs. If we compare these results to those in the
top panel, we find the same ordering of DIBs -- i.e., W($\lambda$5797)
still has the strongest PC$_1$ projection and W($\lambda$5769) still has
the weakest. This confirms that PC$_1$ traces the amount of
DIB-producing material in the top panel, as well. This conclusion is
further established by the bottom left panel, illustrating the PCA
results when only the line of sight parameters are considered. Here,
PC$_1$ appears to be quite different: Both N(H$_2$) and E(B-V) have
stronger PC$_1$ projections than N(H), whereas in the top panel, the
N(H) and E(B-V) projections on PC$_1$ are comparable, while that of
N(H$_2$) is much shorter. This means that PC$_1$ is tracing something
slightly different in each case. In the former, it seems to trace
something more closely related to the amount of dust (given the large
E(B-V) projection), whereas in the latter, the amount of gas plays a
stronger role. Since the DIB carriers are widely accepted as begin
gaseous species, it seems reasonable that N(H) would more strongly
influence the strength of DIBs than E(B-V) as illustrated in our full
PCA results, so this too is consistent with our interpretation that
PC$_1$ traces the amount of DIB-producing material. The fact that
these cases are so different while the just-DIB results are nearly
identical to those obtained using the full data set implies that our
results are driven by the DIBs, rather than the line of sight
parameters.

\begin{figure}
\begin{center}
\includegraphics[width=8.5cm,  trim={1cm, 6.5cm, 1cm, 7cm},clip]{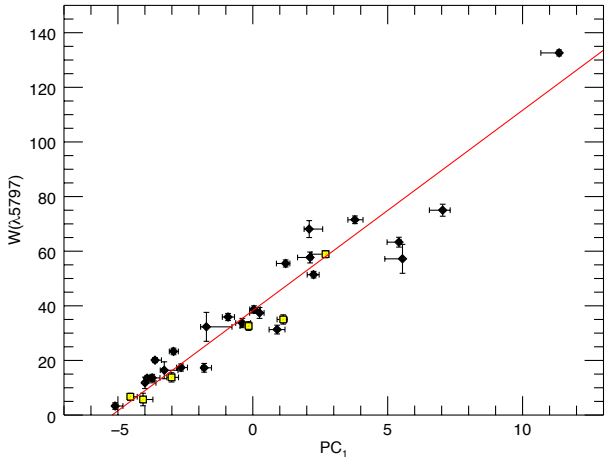}
\caption{\label{fig:PC1_5797} W($\lambda$5797) and PC$_1$ have a very
  strong correlation ($r$=0.957). The equation of the best-fit line is
  $W(\lambda5797)=7.31(PC_1)+38.25$.  Be stars are indicated by yellow
  squares. }
\end{center}
\end{figure}

Interestingly, the largest projection onto PC$_1$ comes from the
$\lambda$5797 DIB, meaning that PC$_1$ has the strongest correlation
with W($\lambda$5797) ($r$ =0.957 -- see
Figure~\ref{fig:PC1_5797}). This result also carries over to the
just-DIB case ($r$=0.963 -- not shown). Hence, the best measure for
the amount of DIB-producing material in the line of sight is
W($\lambda$5797) and not N(H) or E(B-V). Therefore, W($\lambda$5797)
is conceivably a more appropriate normalization factor for DIB
strengths than E(B-V). This is discussed more in
Section~\ref{sec:discussion}.

After establishing that PC$_1$ traces the amount of DIB-producing
material along the line of sight, we can define a new parameter,
N$_{\text{DIB}}$, describing this value. The transformed data points
derived from PCA are, by definition, centered at zero and therefore
approximately half of our sightlines have negative PC$_1$ values. For
N$_{\text{DIB}}$ to be physically meaningful, we need to have only
positive values. Since our PCs were calculated from standardized
variables, we have a way of adjusting our values. Using equation
\ref{eqn:standardization}, we subtracted the mean divided by the
standard deviation from each variable prior to performing PCA. To
correct for this, we must add these values to each term in our PC$_1$
equation (see Equation \ref{eqn:PC1_eqn}). Thus, we calculate a
constant, $C_1$, from the weighted fractions of $\bar{x_i}/s_x{_i}$
for each variable, $x_i$. Explicitly,

\begin{equation}
C_1 = \sum_{i} \frac{a_{i,1}(\bar{x_i})}{s_{x_i}}
\end{equation}

and thus, PC$_1$ can be expressed in its uncentered form:

\begin{align}
PC_{1, uncentered} &= PC_1 + C_1 \\
&= PC_1 + 6.616 \nonumber
\end{align}

It is also worth noting that, since the PCs define directions in
space, the scaling is arbitrary. We can use the tight correlation with
W($\lambda$5797) to rescale PC$_1$ such that as the strength of
$\lambda$5797 doubles, so too does N$_{\text{DIB}}$. The result is
Equation~\ref{eqn:Ndib}.

\begin{align}
\label{eqn:Ndib}
N_{DIB}&=0.136(PC_{1, uncentered}) \\
&=0.136(PC_1) + 0.900 \nonumber
\end{align}

Since PC$_1$ is not directly measurable, we provide an alternate
expression in terms of W($\lambda$5797).

\begin{equation}
\label{eqn:Ndib_5797}
N_{DIB} \approx 0.0185W(\lambda5797)+0.19
\end{equation}

The constant in Equation~\ref{eqn:Ndib_5797} indicates that some DIB
carriers are present before the $\lambda$5797 carriers start to
form. This result is further discussed in Section
\ref{sec:discussion}.

\subsection{PC$_2$: strength of the ambient radiation field}

PC$_2$ represents the second largest source of variations in the DIB
strengths and line of sight parameters in our sample, and must be a
parameter that is uncorrelated with PC$_1$. The largest contribution
to PC$_2$ comes from W($\lambda$5797)/W($\lambda$5780) (see
Fig.~\ref{fig:PC1-PC2} and Eq.~(\ref{eqn:PC2_eqn})) which correlates
strongly (but negatively) with PC$_2$. The second largest projection
onto PC$_2$ is W($\lambda$6284), in the positive direction.  Note that
some of our original parameters are hardly influenced by PC$_2$, most
notably W($\lambda$5797), W($\lambda$6379), N(H) and E(B-V). 

These correlations (and lack thereof) are valuable clues to infer the
physical quantity that is represented by PC$_2$. We start by noting
that W($\lambda$5797)/W($\lambda$5780) is a variable that is clearly
related to physical conditions, and has in particular been linked to
the amount of exposure to the ambient UV radiation field \citep[see
  e.g.][]{Krelowski&Sneden, Herbig1995, Cami1997, Sonnentrucker1997,
  Krel1999, Vuong-Foing, Cox:LMC-DIBs, Friedman2011,
  2011A&A...533A.129V}. More specifically, these studies have shown
that the $\lambda$5780 DIB is weak in shielded environments, but
becomes significantly stronger with increasing UV exposure in diffuse
clouds; the $\lambda$5797 DIB on the other hand is already strong in
shielded environments, and is largely indifferent to UV exposure in
most diffuse clouds. In the harshest environments (e.g. Orion), both
DIBs are very weak or absent. This different dependence on the
radiation field then results in the more sheltered $\zeta$-type clouds
having a high W($\lambda$5797)/W($\lambda$5780) ratio and the more
exposed $\sigma$-type cloud environments having a low ratio. This
dependence on the radiation field is also responsible for the
so-called ``skin effect'' -- the fact that DIB carriers seem to be
concentrated near the surfaces of clouds \citep[see
  e.g.][]{1991MNRAS.252..234A, Herbig1995, Cami1997,
  2011A&A...533A.129V}.

One way such behaviour can be understood is by considering that the
DIB carriers are molecules in a specific ionization state, and the
changing DIB strengths reflect the molecules undergoing ionization
with the carrier of the $\lambda$5797 having a lower ionization
potential than the $\lambda$5780 DIB carrier
\citep{Cami1997,Sonnentrucker1997}. Note that the charge balance is
really determined by the interplay between ionization and
recombination, and thus not only depends on the radiation field, but
also on the density. Similarly, the DIB behaviour can be explained by
the DIB carriers being a specific hydrogenation or protonation state
\citep{Vuong-Foing, LePage:hydrogenation, LePage:hydrogentationII};
also in this case, the density plays a role. 

Thus, the W($\lambda$5797)/W($\lambda$5780) ratio certainly traces
cloud depth, which encompasses UV exposure, density, and possibly also
temperature. Given the strong correlation with PC$_2$, PC$_2$ could
thus trace directly the strength of the ambient radiation field
$G_0$\footnote{$G_0$ is a convenient measure for the strength of the
  FUV field that is often used in the context of Photo-Dissociation
  Regions \citep[PDRs;][]{TielensHollenbach:PDRs}. It is the FUV field
  measured in units of the equivalent \citet{Habing:IS_FUV} flux of
  1.6$\times$10$^{-3}$ ergs~cm$^{-2}$~s$^{-1}$ appropriate to the
  average interstellar medium.}, or alternatively it could be a
measure for $G_0/n_e$ and reflect the balance between ionization and
recombination (or similarly between hydrogenation and
de-hydrogenation). 

We dismiss this second interpretation for two reasons. First,
\cite{dibs_ne} showed that several of the DIBs included in our sample
are independent of n$_e$, contrary to the large variations we see
among their PC$_2$ values. Furthermore, \cite{Kos&Zwitter} found no
differences in n$_e$ values between $\sigma$ and $\zeta$ sightlines,
despite W($\lambda$5797)/W($\lambda$5780) having the largest projection on
PC$_2$. Thus, we conclude that PC$_2$ must be tracing changes in G$_0$
only, although of course this is in turn can be determined by the
cloud depth. 

If we dichotomize sightlines into $\sigma$ and $\zeta$ clouds, this
becomes more clear. Referring again to Figure~\ref{fig:PC1-PC2}, the
negative projections onto PC$_2$ are all indicators of a sheltered
$\zeta$ environment: We see a strong
W($\lambda$5797)/W($\lambda$5780), hydrogen in molecular form
(i.e. large f(H$_2$)), and the existence of C$_2$ \citep[demonstrated
  by the four C$_2$-DIBs -- $\lambda$5769, $\lambda$5546,
  $\lambda$5513, and $\lambda$4964; ][]{Thorburn}. In contrast,
positive PC$_2$-projections are consistent with more exposed
conditions: we observe hydrogen in neutral form, as well as a strong
projection from $\lambda\lambda$5780, 4428 and 6284 -- DIBs known to
correlate well with \ion{H}{1} \citep{Herbig1993}. Finally, both N(H)
and E(B-V) contribute almost nothing to PC$_2$. Again, this is
consistent with our interpretation since both variables contribute to
the amount of material but are generally unaffected by changes in
G$_0$. Note that $\lambda$5797 and $\lambda$6379 exhibit the same
behavior, as was found previously \citep{Cami1997, Kos&Zwitter,
  Lan:SDSS_DIBs}. The case of $\lambda$5797 is especially
interesting. Although we use W($\lambda$5797)/W($\lambda$5780) to
differentiate between regions of high and low UV exposure, these
changes almost exclusively reflect the $\lambda$5780 carrier. Thus,
while the carrier of $\lambda$5780 is sensitive to changes in G$_0$,
$\lambda$5797 survives at a steady strength per unit reddening over a
wide range of conditions. This indifference to physical conditions is
essentially why this DIB is a good tracer for the amount of DIB
carrier material.

Comparing the three separate panels of Figure~\ref{fig:PC1-PC2}, we
once again see that the PCA results from the line of sight parameters
(bottom left) are quite different. N(H), in particular, has a strong,
positive projection on PC$_2$, making it clear that PC$_2$ traces
something other than G$_0$ in this case. This is not surprising since
PC$_2$ necessarily depends on PC$_1$, and PC$_1$ was different between
the full PCA and the PCA for the line of sight parameters only. The
PCA results using only the DIBs, however, are nearly identical to
those obtained with all 23 variables. This suggests that changes in
PC$_2$ -- that is to say, changes in G$_0$ -- are directly observable
through the relative strengths of DIBs.

For the sake of clarity, we further formalize our association of
PC$_2$ with the radiation field by defining a new parameter
G$_{\text{DIB}}$. Ideally, we would like to scale this parameter to
represent actual $G_0$ values; however, without knowledge of specific
G$_0$ values for our targets, we cannot do this. Therefore, we leave
PC$_2$ as it is, and simply assign the name G$_{\text{DIB}}$:
\begin{equation}
\label{eqn:Gdib_pc2}
G_{DIB} = PC_2
\end{equation}

\begin{figure}
\begin{center}
\includegraphics[width=8.5cm,  trim={1cm, 6.5cm, 1cm, 7cm},clip]{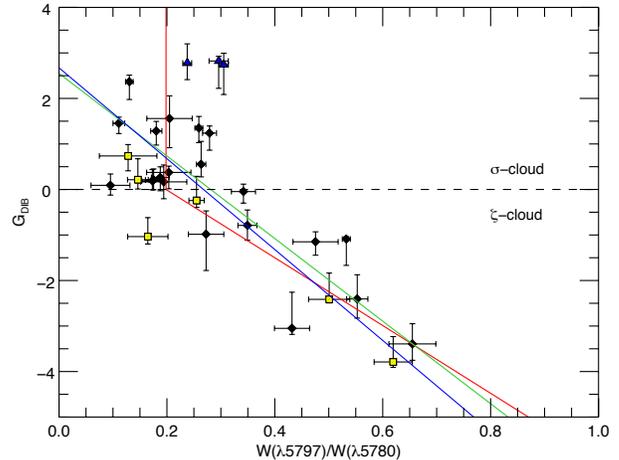}
\caption{\label{fig:PC2_57975780} G$_{\text{DIB}}$ as a function of
  W($\lambda$5797)/W($\lambda$5780). Be stars are indicated by yellow
  squares; three possible outliers (HD~15137, HD~198478, and
  HD~209975; located at G$_{\text{DIB}}\approx 3$) are shown as blue
  triangles. We show the three fits discussed in the text:
  Eq.~(\ref{Gdib_linear}) in green; the fit without outliers from
  Eq.~(\ref{eqn:Gdib_wo_outliers}) in blue and the piece-wise function
  from Eq.~(\ref{eqn:Gdib_57975780}) in red. See text for details. }
\end{center}
\end{figure}

Since W($\lambda$5797)/W($\lambda$5780) correlates most strongly with
PC$_2$, we will approximate G$_{\text{DIB}}$ as a function of
W($\lambda$5797)/W($\lambda$5780). Figure~\ref{fig:PC2_57975780} shows
G$_{\text{DIB}}$ as a function of the W($\lambda$5797) to
W($\lambda$5780) strength ratio. Fitting a straight line to the data
gives the following relation with $r=-0.716$ (shown in green in
Fig.~\ref{fig:PC2_57975780}):

\begin{equation}
\label{Gdib_linear}
G_{DIB} \approx -9.07[W(\lambda5797)/W(\lambda5780)] + 2.55
\end{equation}

Arguably, a linear fit is not appropriate since there are three
sightlines that strongly deviate from this relation -- HD~15137,
HD~198478, and HD~209975. However, the interstellar features for these
sightlines are quite broad, suggesting that there may be more than one
cloud in the line of sight that we are unable to resolve. If this is
the case, then we cannot expect them to follow the same trend as true
single clouds. Ignoring these three sightlines in our linear fit, we
obtain (with $r=-0.832$; shown in blue in
Fig.~\ref{fig:PC2_57975780}):

\begin{equation}
\label{eqn:Gdib_wo_outliers}
G_{DIB}=-9.98[W(\lambda5797)/W(\lambda5780)]+2.67
\end{equation}

Alternatively, these sightlines are in fact single clouds and
represent a true trend in the data. With this interpretation, the
linear dependence breaks down for positive G$_{\text{DIB}}$ values
(i.e, $\sigma$-type clouds) and G$_{\text{DIB}}$ increases
independently of W($\lambda$5797)/W($\lambda$5780). As previously discussed,
W($\lambda$5797) is independent of PC$_2$, while W($\lambda$5780) shows a
strong dependence; hence, this trend is most easily explained as an
ionization effect of the $\lambda$5780 carrier. We propose that
carrier of $\lambda$5780 is a cation or a dehydrogenated species
\citep[see e.g.][]{Cami1997, Sonnentrucker1997}. As G$_0$ increases,
more of these molecules become ionized/dehydrogenated so W($\lambda$5780)
becomes stronger. Finally, when a certain G$_0$ is reached, we enter
the $\sigma$-cloud regime, where the cation/dehydrogenated state
dominates. Hence, W($\lambda$5780) is no longer dependent on the
environment, and $\lambda$5797 and $\lambda$5780 exist in
approximately equal proportions. We can model this using a piece-wise
function. For $\sigma$ clouds, W($\lambda$5797)/W($\lambda$5780) has a mean
value of 0.198$\pm$0.064, with a maximum W($\lambda$5797)/W($\lambda$5780)
value around 0.3. For $\zeta$ clouds, the linear relation holds. We
fit a line to the negative G$_{\text{DIB}}$ values, forcing the
intercept through the $\sigma$ cloud mean using the
MPFITEXY\footnote{This is a functional linear regression taking into
  account errors in both $x$ and $y$ values.}  routine
\citep{MPFITEXY} based on the MPFIT package \citep{MPFIT}. We obtain a
correlation coefficient of $-$0.75 for $\zeta$ clouds. The resulting
equation is:

\begin{equation}
G_{DIB} \approx
\begin{cases}
\label{eqn:Gdib_57975780}
\text{undefined}, & \text{if } \frac{W(\lambda5797)}{W(\lambda5780)} \lesssim 0.3 \\
-8.44(\frac{W(\lambda5797)}{W(\lambda5780)}) +1.94, & \text{if } \frac{W(\lambda5797)}{W(\lambda5780)} \gtrsim 0.3
\end{cases}
\end{equation}

From this, it is clear that W($\lambda$5797)/W($\lambda$5780) is a suitable
measure for G$_0$ within $\zeta$ clouds, whereas for $\sigma$ clouds,
it may or may not be appropriate.

\subsection{PC$_3$}

\begin{figure}[h]
\begin{center}$
\begin{array}{crr}
 \includegraphics[width=80mm,trim={1.5cm, 6.1cm, 1.5cm, 6.1cm}]{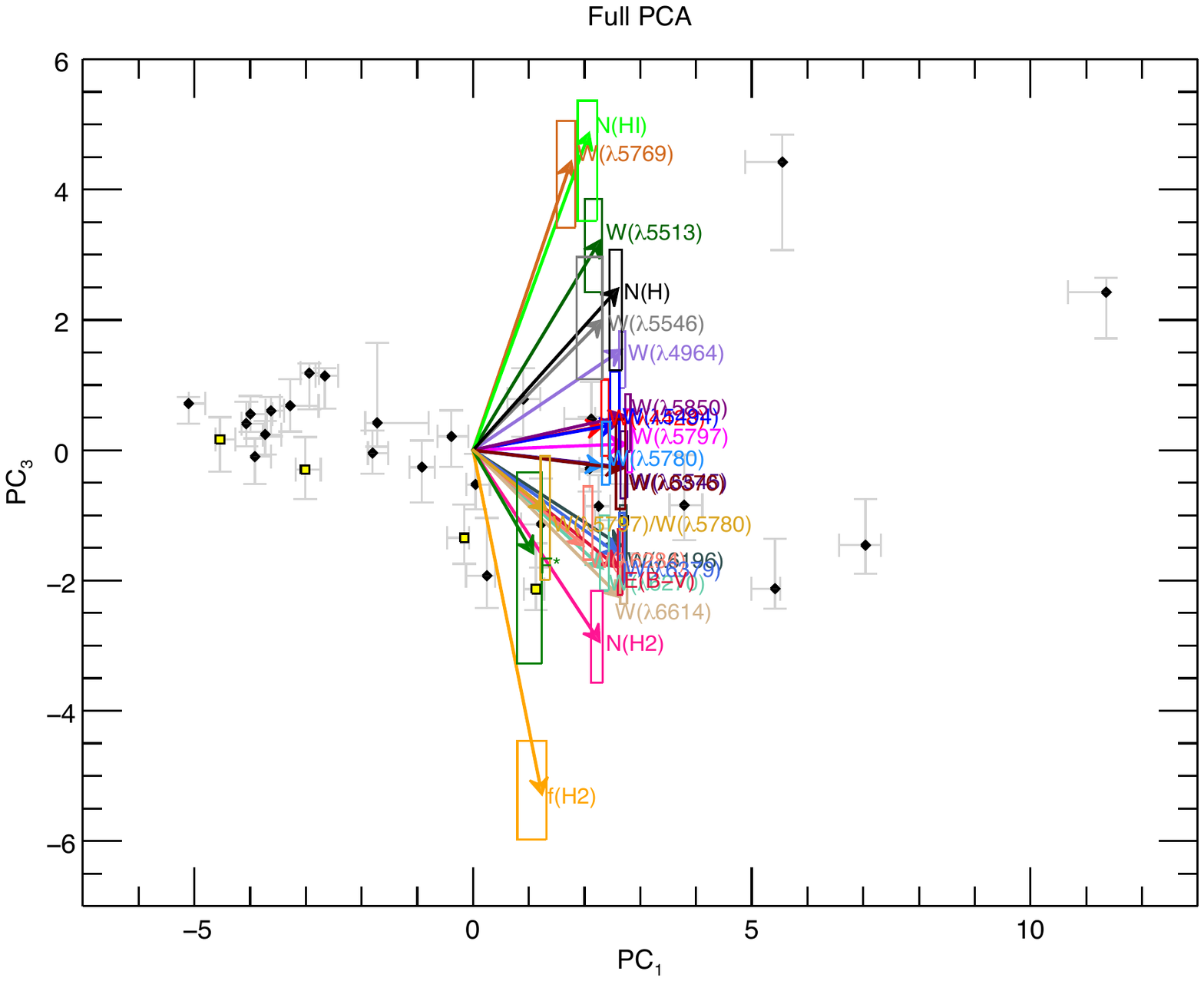}\\
 \includegraphics[width=80mm,trim={1.5cm, 6.1cm, 1.5cm, 6.1cm}]{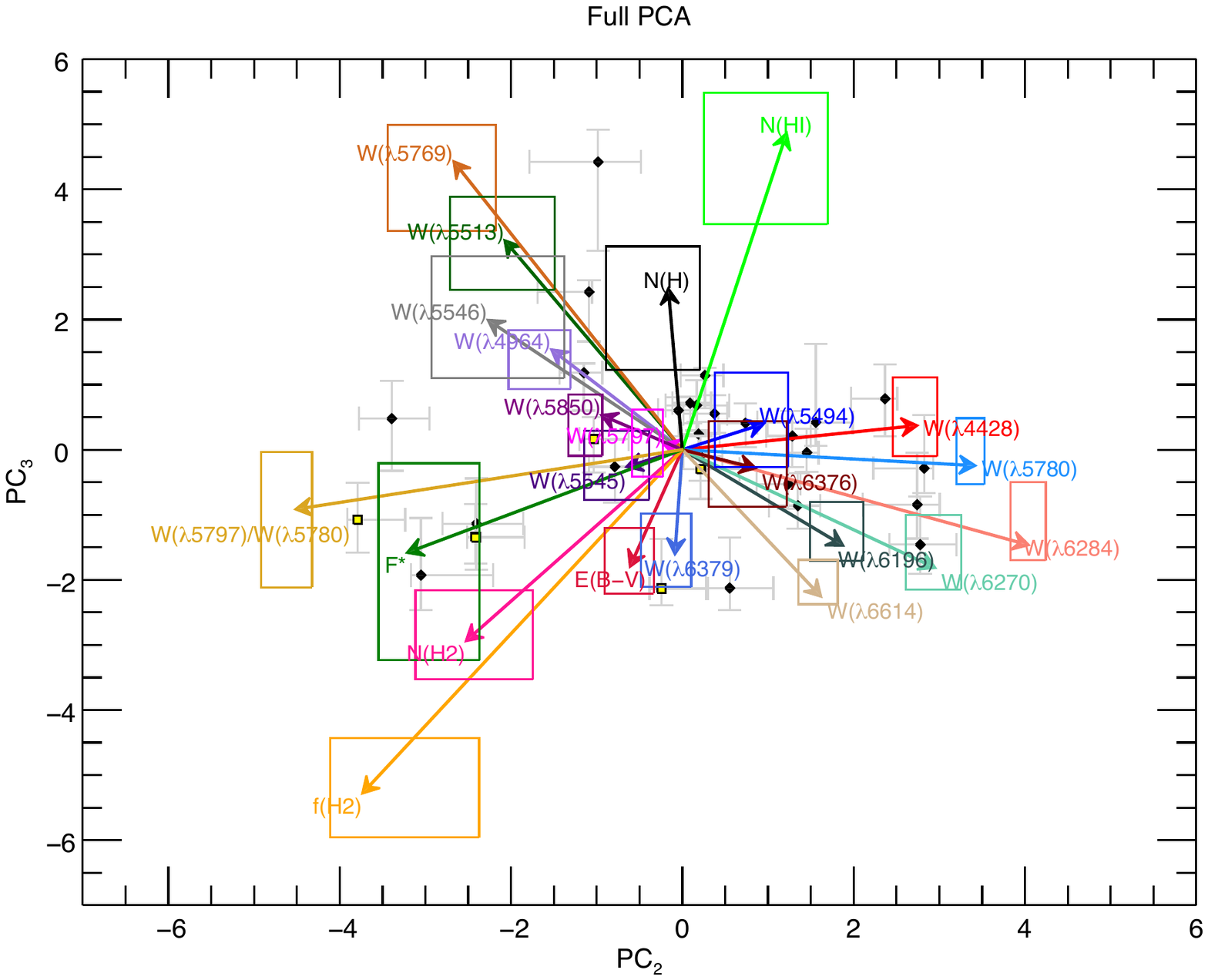}\\
\end{array}$    
\caption{\label{fig:PC3} (\textit{Top}:) PC$_1$-PC$_3$ biplot.
  (\textit{Bottom}:) PC$_2$-PC$_3$ biplot. Be stars are indicated by
  yellow squares. The outlying data point is HD~147933 with a PC$_3$
  value of 4.42.} 
\end{center}
\end{figure}

Figure~\ref{fig:PC3} shows the PC$_1$-PC$_3$ and PC$_2$-PC$_3$
biplots; note that we no longer consider the PCA results from just
DIBs as the uncertainties are too large to properly interpret the
results. Instead, we focus only on the full PCA results.

Because PC$_3$ is necessarily uncorrelated with PC$_1$ and PC$_2$, it
must trace something that is fairly independent of both the amount of
material and the ambient UV field. Some clues about PC$_3$ are
revealed in Figure~\ref{fig:PC3}. For negative PC$_3$ values, we see
strong projections from f(H$_2$) and N(H$_2$). For positive values, we
see strong projections from N(\ion{H}{1}) and the C$_2$-DIBs. We may
therefore interpret PC$_3$ as tracing some difference between the
amount of H$_2$ and the amount of both C$_2$ and
\ion{H}{1}. \cite{VD&B} showed that C$_2$ tends to form in cooler,
denser regions than H$_2$ so one could interpret PC$_3$ to trace the
temperature and density. Note that in that interpretation, colder and
denser environments correspond to higher values for PC$_3$. We test
this hypothesis by considering the projections on PC$_3$. Although the
vector for F$_\star$ has large uncertainties, the projection is
clearly negative. This would then imply that more depletion occurs in
warmer, more diffuse regions. This result seems counterintuitive, and
it is therefore unlikely that PC$_3$ traces temperature and density.

A careful analysis of the biplots reveals more clues about
PC$_3$. There is one clear outlier, HD~147933, which has a PC$_3$
value of 4.42. PC$_3$ therefore traces some characteristic that sets
HD~147933 apart from the other lines of sight included in this
study. This target, found within the $\rho$ Ophiuchi dark cloud, has
been well-studied. A notable characteristic is its unusually large
dust grains \citep{pOphA}, resulting in a ratio of hydrogen to color
excess 2.7 times greater than the mean interstellar value
\citep{BSD1978}. Two plausible explanations are therefore grain size
or the gas-to-dust ratio.

We first consider the interpretation that PC$_3$ traces grain size,
with positive PC$_3$ projections corresponding to larger grains and
negative projections corresponding to smaller grains. Since small dust
grains dominate the total dust grain surface area, and the only
effective mechanism for producing interstellar H$_2$ is on the surface
of dust grains \citep{GS63}, it makes sense that the strongest
negative PC$_3$ projections come from f(H$_2$) and N(H$_2$). Moreover,
E(B-V) has a negative projection since small grains dominate reddening
as well. Finally, F$_\star$ has a negative PC$_3$ projection since a
larger surface area of dust grains implies that there is more surface
onto which elements may deplete. In terms of positive projections, we
see a large projection from N(\ion{H}{1}). This too is consistent
since, with larger dust grains and a reduced surface area, \ion{H}{1}
cannot effectively be converted into H$_2$.

\begin{figure}
\begin{center}
\includegraphics[width=8.5cm,  trim={1cm, 7cm, 1cm, 7cm}, clip]{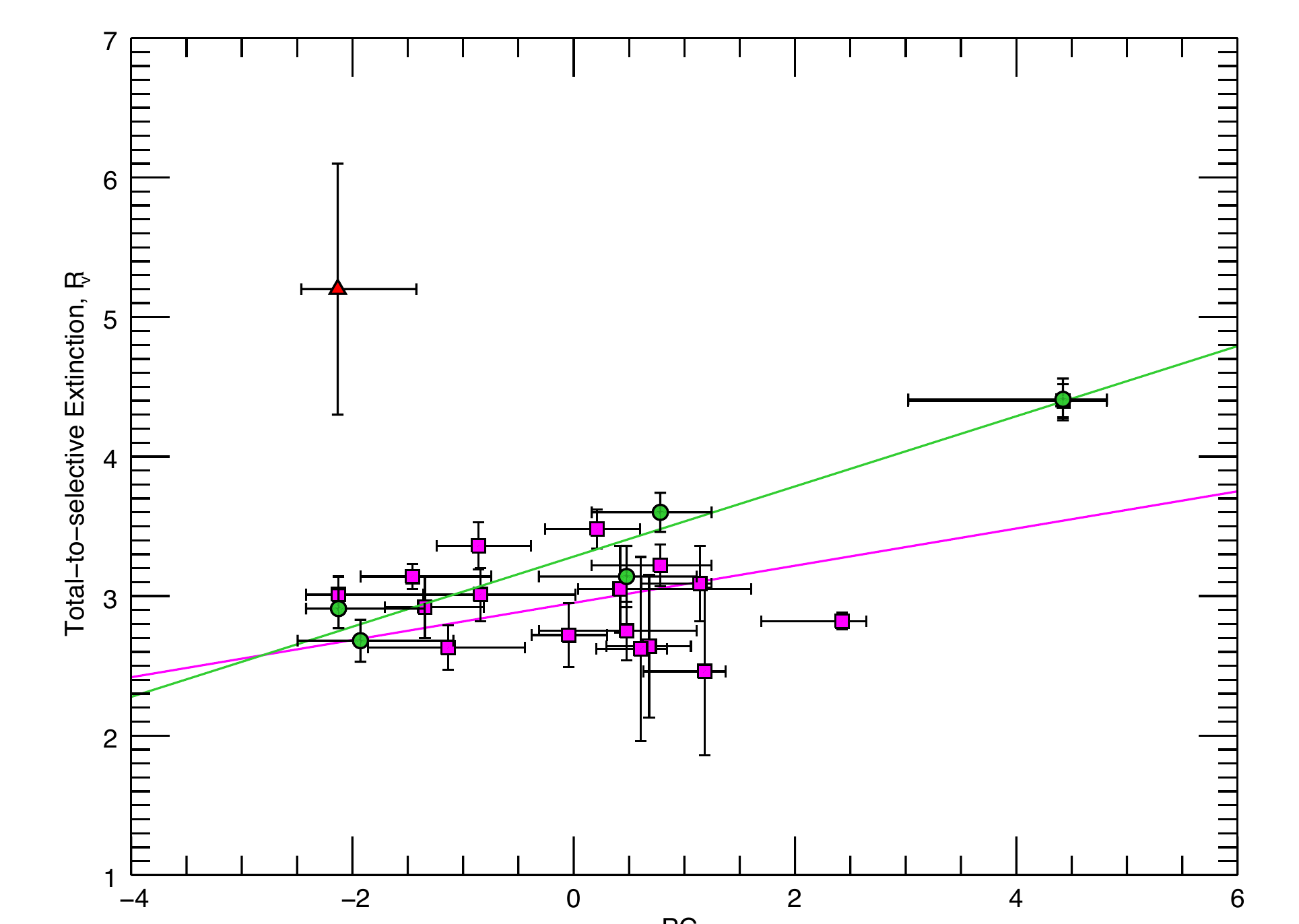}
\caption{\label{fig:PC3_Rv} Total-to-selective extinction ratio,
  R$_v$, as a function of PC$_3$. R$_v$ values are taken from:
  \citet[][magenta squares]{Wegner2003}; \citet[][green
    circles]{Fitz&Massa2007} and \citet[][red
    triangles]{Andersson&Potter2007}. Note that there is some overlap
  between the \cite{Wegner2003} and \cite{Fitz&Massa2007} targets. The
  magenta and green lines are straight line fits (using IDL's FITEXY
  routine) to the corresponding data sets by \citet{Wegner2003} and
  \citet{Fitz&Massa2007} respectively. We did not merge both data sets
  because of some discrepancies between the reported R$_v$ values in
  both sets. Note that the value found for HD~110432 by
  \citet{Andersson&Potter2007} deviates significantly from the mean
  trend.}
\end{center}
\end{figure}

To further test this idea, we plot the total-to-selective extinction,
R$_v$, as a function of PC$_3$ in Figure~\ref{fig:PC3_Rv}. R$_v$ tends
to increase with grain size, so if our interpretation is correct, we
would expect to see R$_v$ increasing with PC$_3$. Although a weak
trend is apparent, HD~110432 (shown as a red triangle) is hard to
reconcile with the rest of the data. Part of this discrepancy may be
due to the fact that HD~110432 is a Be star. \cite{Rachford2001}
corrected for the CS contribution and found that an R$_v$ value of 3.3
is more appropriate, although this still implies an above-average
grain size.

Alternatively, PC$_3$ may trace the gas-to-dust ratio. With this
interpretation, positive projections indicate larger gas-to-dust
ratios while negative projections indicate larger dust-to-gas
ratios. Both f(H$_2$) and N(H$_2$) have large, negative projections,
which again is consistent with a larger dust grain surface area and
therefore a higher dust-to-gas ratio. E(B-V) also increases in this
direction, with similar reasoning. Finally, we see a large projection
from F$_{\star}$. Noting that the large grains toward HD~147933 come
from grain coagulation rather than elemental depletions
\citep{Jura1980}, this too is consistent: If there is more dust, then
there is more material onto which gas may deplete. Both N(H) and
N(\ion{H}{1}) have positive projections, consistent with higher ratios
of gas. N(\ion{H}{1}), in particular, has a large projection since a
large quantity of dust is required to transform \ion{H}{1} into H$_2$.

\begin{figure}
\begin{center}
\includegraphics[width=8.5cm,  trim={1cm, 7cm, 1cm, 7cm},clip]{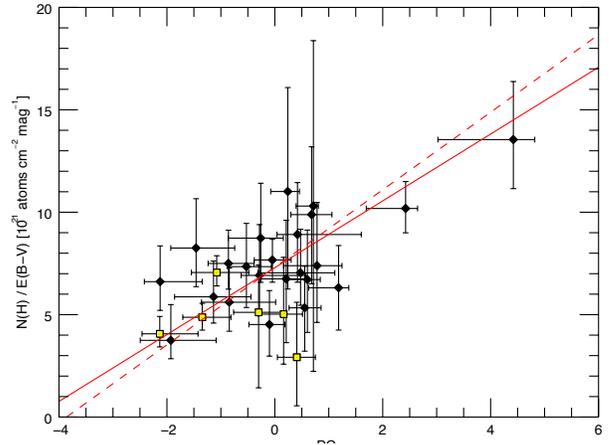}
\caption{\label{fig:PC3_dtg}The total column of hydrogen per unit
  reddening as a function of PC$_3$. Yellow squares are used to
  indicate Be stars, which mostly fall below the trend line. The solid
  red line represents a straight line fit through all data points
  (Eq.~\ref{Eq:NH_EBV_PC3_1}; the dashed red line is the same
  excluding the two outliers (Eq.~\ref{Eq:NH_EBV_PC3_2}). Note that
  HD~143275 is not included in this plot since it has an E(B-V) value
  of zero.}
\end{center}
\end{figure}

Figure~\ref{fig:PC3_dtg} shows the ratio of hydrogen to color excess,
N(H)/E(B-V), plotted against PC$_3$. The general trend is apparent,
and can be represented by (using IDL's FITEXY routine):
\begin{equation}
\label{Eq:NH_EBV_PC3_1}
\frac{\rm N(H)\times 10^{-21}}{E(B-V)} = (7.29\pm0.23) + PC_3 \times (1.63\pm0.26)
\end{equation}
However, the trendline appears to be influenced by the two outliers --
HD~147933 and HD~207198 -- and we therfore performed another fit
excluding these two data points; this yields similar results:
\begin{equation}
\label{Eq:NH_EBV_PC3_2}
\frac{\rm N(H)\times 10^{-21}}{E(B-V)} = (7.31\pm0.27) + PC_3 \times (1.89\pm0.26)
\end{equation}

There is quite a bit of scatter around this trend though. This is
somewhat expected since N(H)/E(B-V) does not measure the gas-to-dust
ratio directly, as E(B-V) measures the degree of reddening but not the
dust mass. Moreover, our values for N(H) do not include any hydrogen
that may exist in ionized form, while both HD~36822 and HD~36861 are
known to lie near \ion{H}{2} regions. Finally, E(B-V) values tend to
be over-estimated for Be stars (see Section \ref{subsec:targets}),
resulting in smaller-than-expected N(H)/E(B-V) values. Be stars are
differentiated in Figure~\ref{fig:PC3_dtg} by square data points.

Although we cannot conclusively state what precisely PC$_3$ is
tracing, it seems clear that it is tracing a property of the dust and
its relation with the gas. We thus favour the gas-to-dust ratio
interpretation here.

Once again, we define a new parameter, GTD, describing the observed
changes along PC$_3$. \cite{BSD1978} showed that the mean interstellar
value of N(H)/E(B-V) is $5.8\times 10^{21}$ atoms cm$^{-2}$
mag$^{-1}$. We rescale PC$_3$ such that GTD gives the ratio of gas to
dust with respect to the mean interstellar value: A GTD value of 1
corresponds to the mean interstellar value of N(H)/E(B-V), and
HD~147933 has a GTD value of 2.7. GTD is therefore expressed as:

\begin{equation}
\label{eqn:GTD}
GTD= 0.318(PC_3) + 1.29
\end{equation}

\subsection{PC$_4$}
\label{subsec:PC4}

\begin{figure}[h]
\begin{center}$
\begin{array}{crr}
           \includegraphics[width=80mm,trim={1.5cm, 6.1cm, 1.5cm, 6.1cm}]{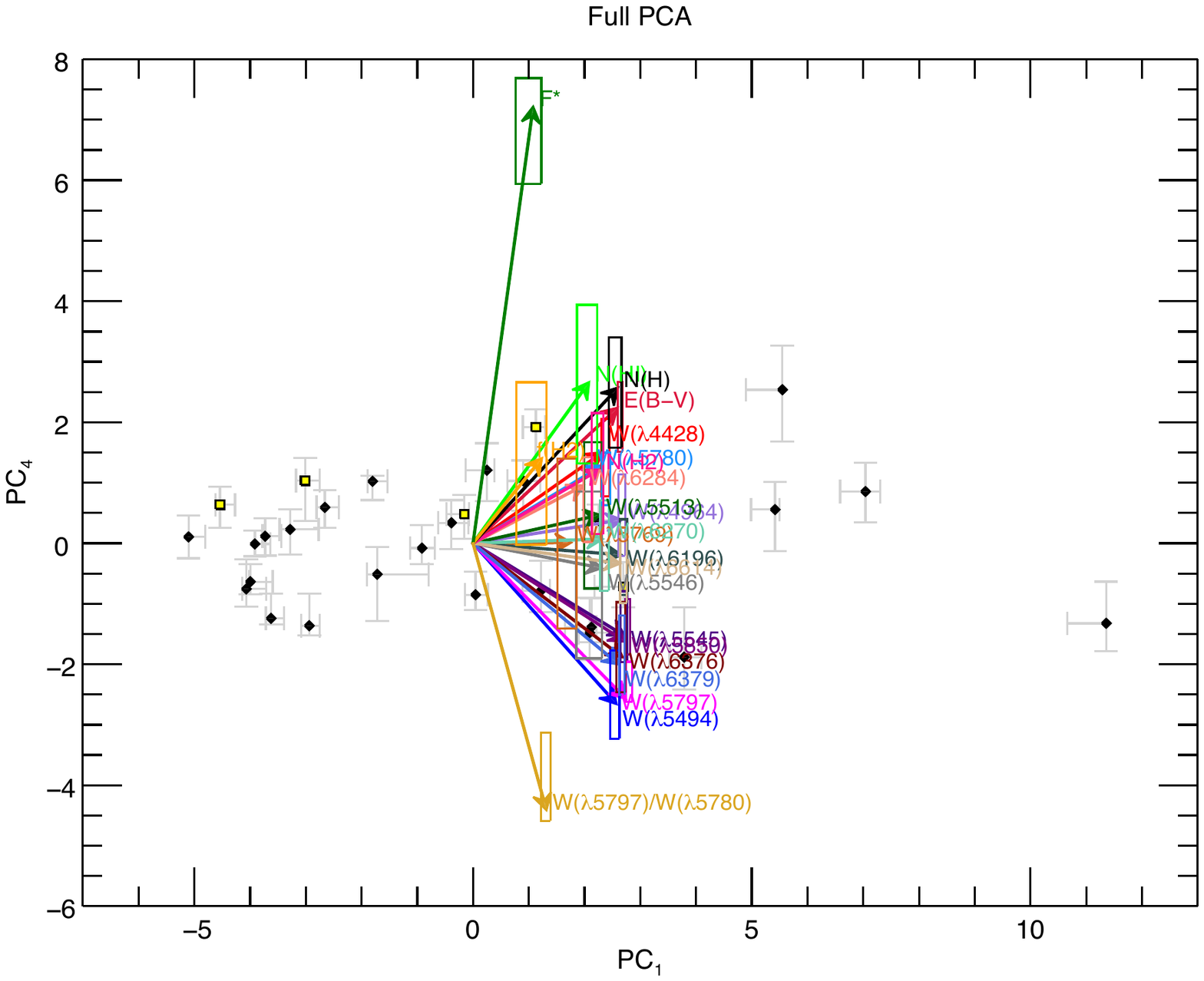}\\
           \includegraphics[width=80mm, trim={1.5cm, 6.1cm, 1.5cm, 6.1cm}]{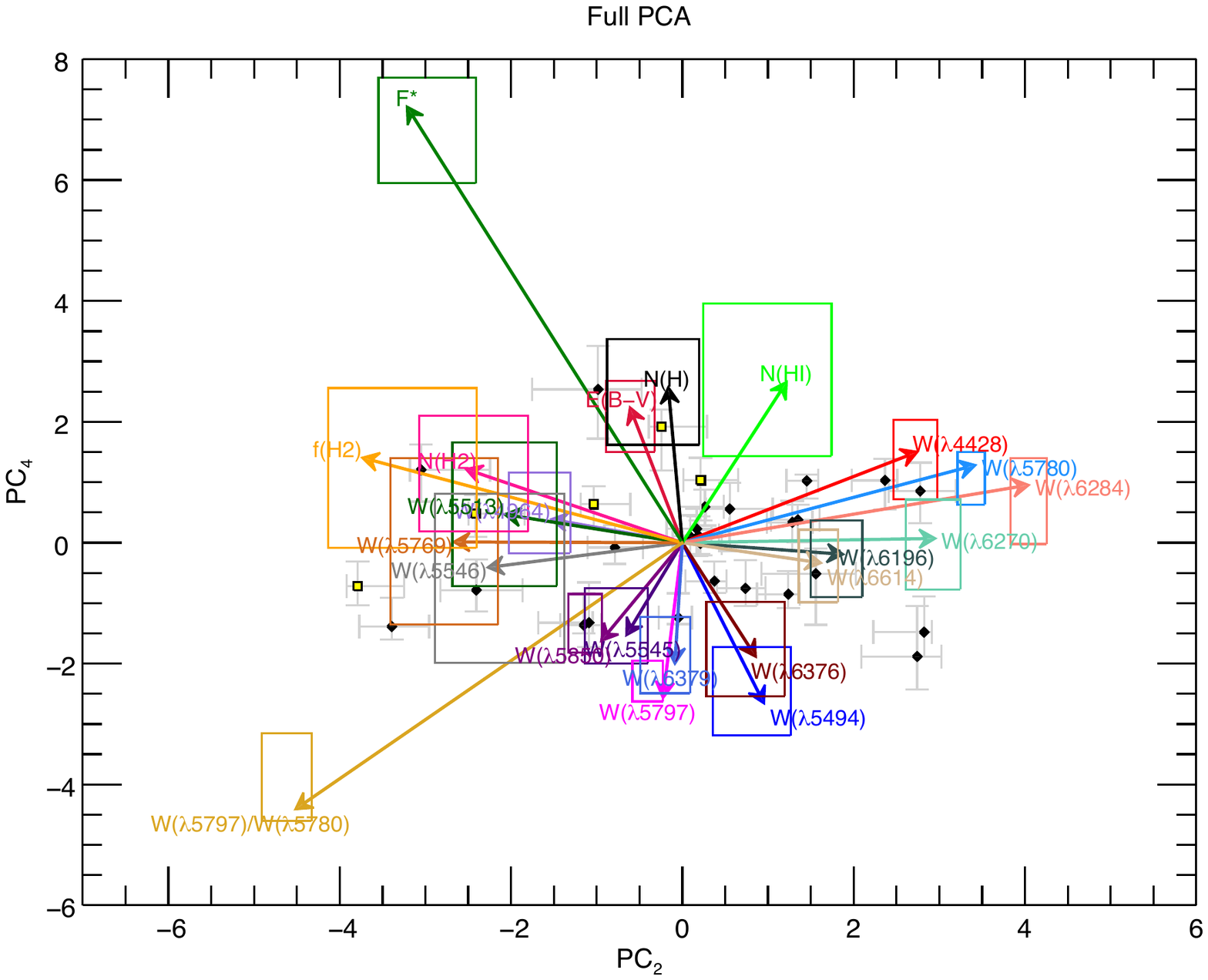}\\
           \includegraphics[width=80mm, trim={1.5cm, 6.1cm, 1.5cm, 6.1cm}]{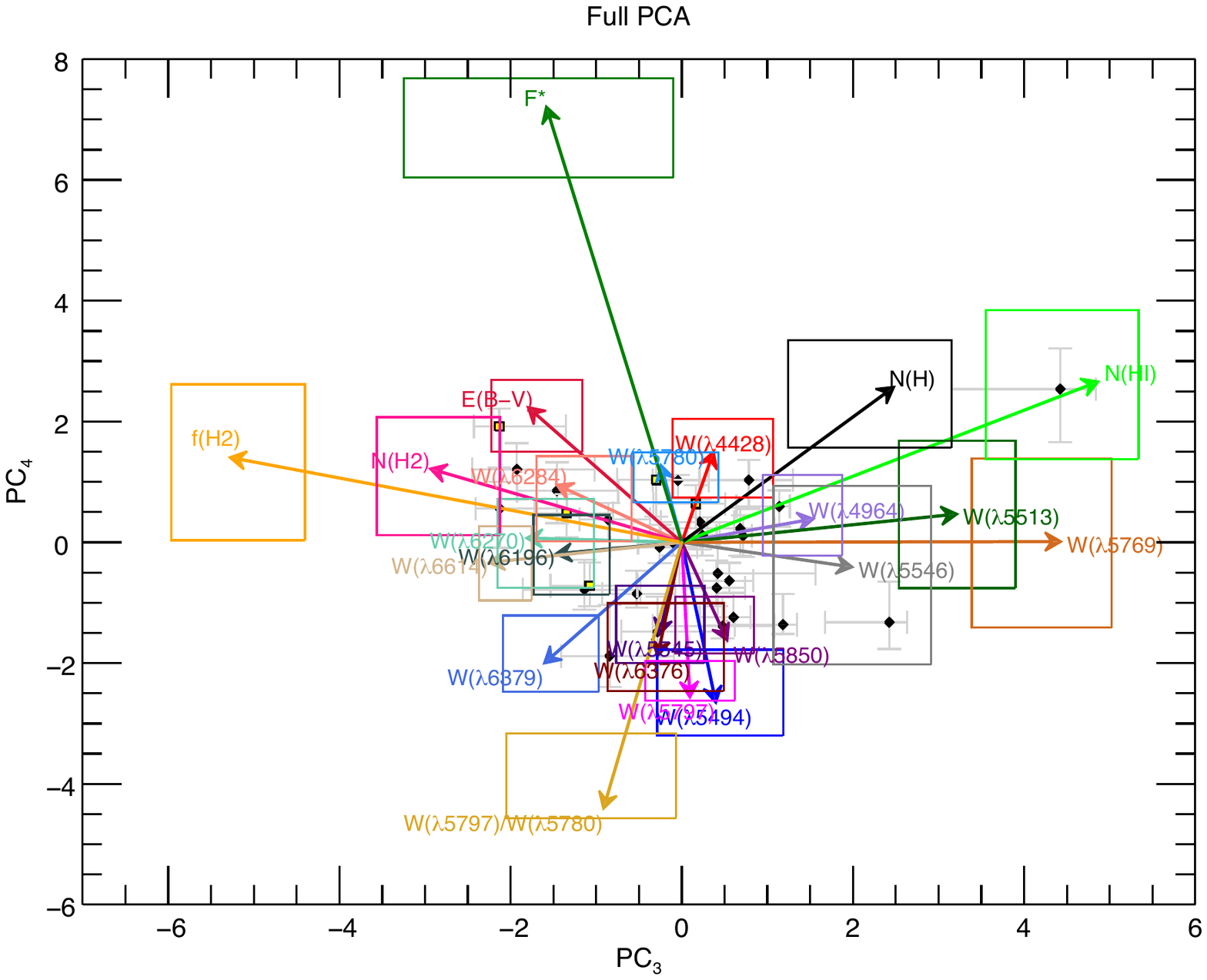}
\end{array}$    
\end{center}
\caption{\label{fig:PC4} (\textit{Top}:) PC$_1$-PC$_4$
  biplot. (\textit{Middle}:) PC$_2$-PC$_4$ biplot. (\textit{Bottom}:)
  PC$_3$-PC$_4$ biplot. Be stars are indicated by yellow squares.} 
\end{figure}

\begin{figure}
\begin{center}
\includegraphics[width=8.5cm,  trim={1cm, 7cm, 1cm, 7cm},clip]{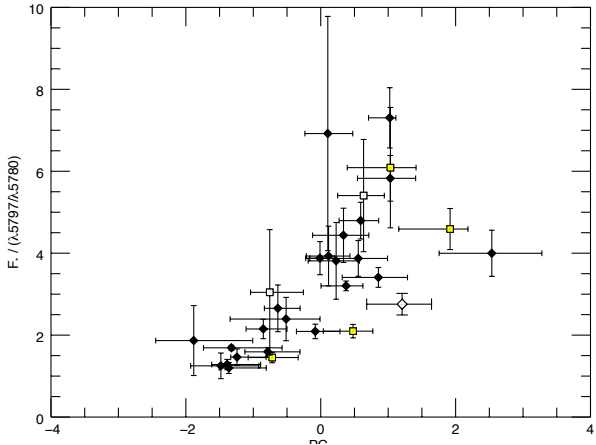}
\caption{\label{fig:PC4_57975780Fstar} PC$_4$ traces some difference
  between F$_\star$ and W($\lambda$5797)/W($\lambda$5780). Here,
  $r$=-0.769. Squares (white and yellow) indicate Be stars. White data
  points (squares and diamonds) indicate synthetically-derived
  F$_\star$ values.} 
\end{center}
\end{figure}

Figure~\ref{fig:PC4} shows the PC$_4$ biplots with PC$_1$ (top),
PC$_2$ (middle), and PC$_3$ (bottom). It is difficult to physically
interpret PC$_4$ due to the large uncertainties on the projections;
however, it is clear that the strongest projections come from
F$_{\star}$ and W($\lambda$5797)/W($\lambda$5780), and thus, PC$_4$ traces
some difference between these two variables. In the previous PCs,
F$_{\star}$ and W($\lambda$5797)/W($\lambda$5780) increased in the same
direction, suggesting that depletions generally increase with cloud
depth. In terms of PC$_4$, however, these two variables increase in
opposite directions. Consequently, a possible interpretation is that
PC$_4$ traces the deviations from the mean relation between these two
variables, although the physical cause for these deviations is
unknown. Figure~\ref{fig:PC4_57975780Fstar} shows the ratio of these
variables plotted as a function of PC$_4$.

\section{Discussion}
\label{sec:discussion}

\begin{figure*}
\begin{center}
\includegraphics[width=18cm, trim={2cm, 5.8cm, 2cm, 6.8cm}] {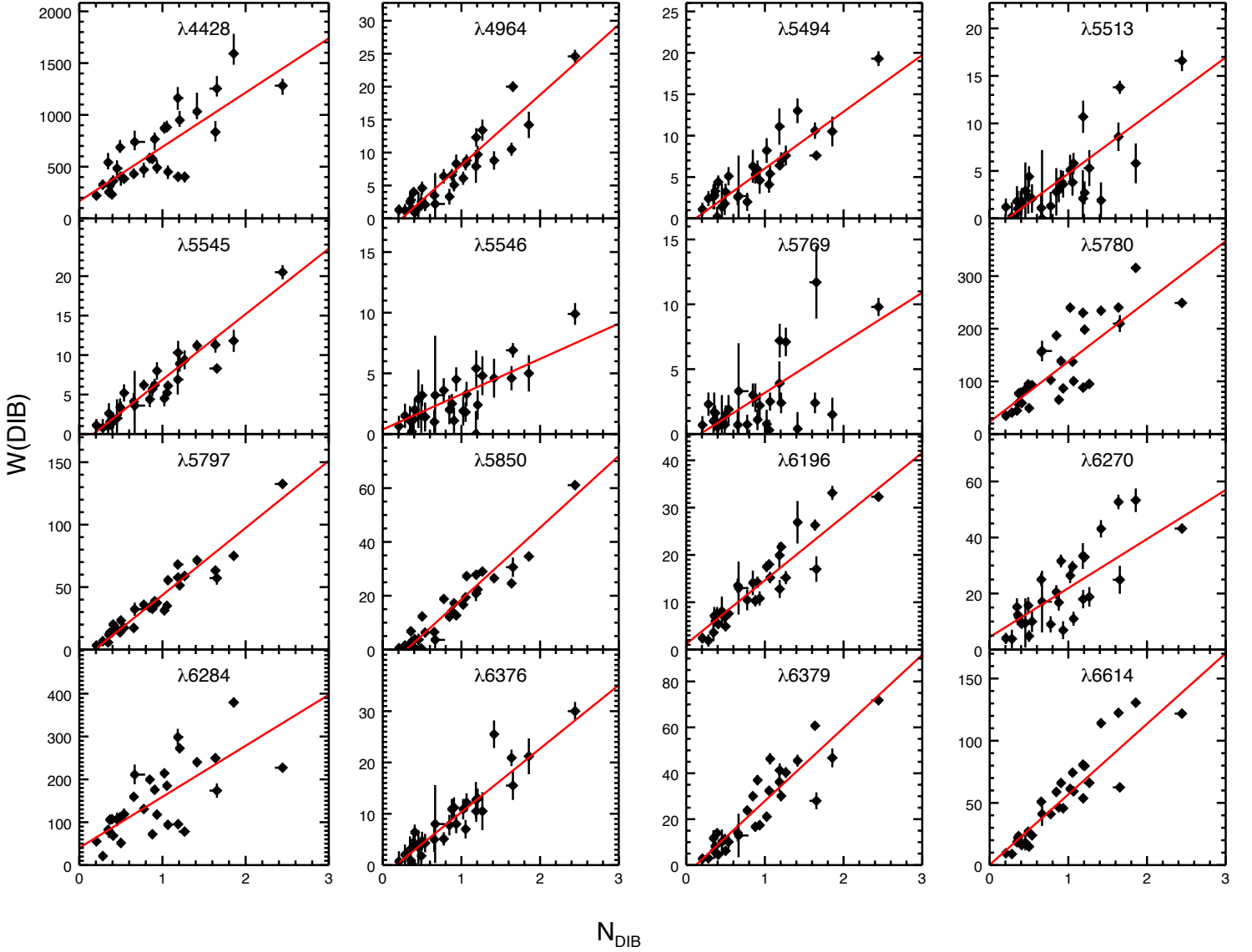}
\caption{\label{fig:dibs_vs_ndib} The strength of each DIB plotted
  against N$_{\text{DIB}}$. In each case, the line of best fit is
  obtained through linear least squares fitting using IDL's FITEXY
  routine. The correlation coefficients, from largest to smallest are
  W($\lambda$5797): 0.957, W($\lambda$5850): 0.952, W($\lambda$5494): 0.948,
  W($\lambda$6376): 0.936, W($\lambda$4964): 0.931 W($\lambda$6196): 0.928,
  W($\lambda$6614): 0.920, W($\lambda$6379): 0.912, W($\lambda$5494): 0.897,
  W($\lambda$6270): 0.811, W($\lambda$4428): 0.811, W($\lambda$5780): 0.807,
  W($\lambda$5546): 0.806, W($\lambda$5513): 0.795, W($\lambda$6284): 0.691,
  and W($\lambda$5769): 0.614. The x-intercepts from smallest to largest
  are: W($\lambda$6284): -0.338$\pm$0.034, W($\lambda$4428):
  -0.317$\pm$0.038, W($\lambda$6270): -0.253$\pm$0.057, W($\lambda$5780):
  -0.202$\pm$0.018, W($\lambda$5546): -0.128$\pm$0.146, W($\lambda$6196):
  -0.087$\pm$0.036, W($\lambda$6614): -0.003$\pm$0.019, W($\lambda$5494):
  0.119$\pm$0.050, W($\lambda$6379): 0.123$\pm$0.024, W($\lambda$6376):
  0.166$\pm$0.059, W($\lambda$5545): 0.171$\pm$0.045, W($\lambda$5769):
  0.177$\pm$0.082, W($\lambda$5797): 0.188$\pm$0.014, W($\lambda$5513):
  0.229$\pm$0.077, W($\lambda$4964): 0.241$\pm$0.040, and W($\lambda$5850):
  0.301$\pm$0.018} 
\end{center}
\end{figure*}

\begin{figure*}[h]
\begin{center}$
\begin{array}{ccc}
           \includegraphics[width=16cm, trim={2cm, 10cm, 1.5cm, 11cm}]{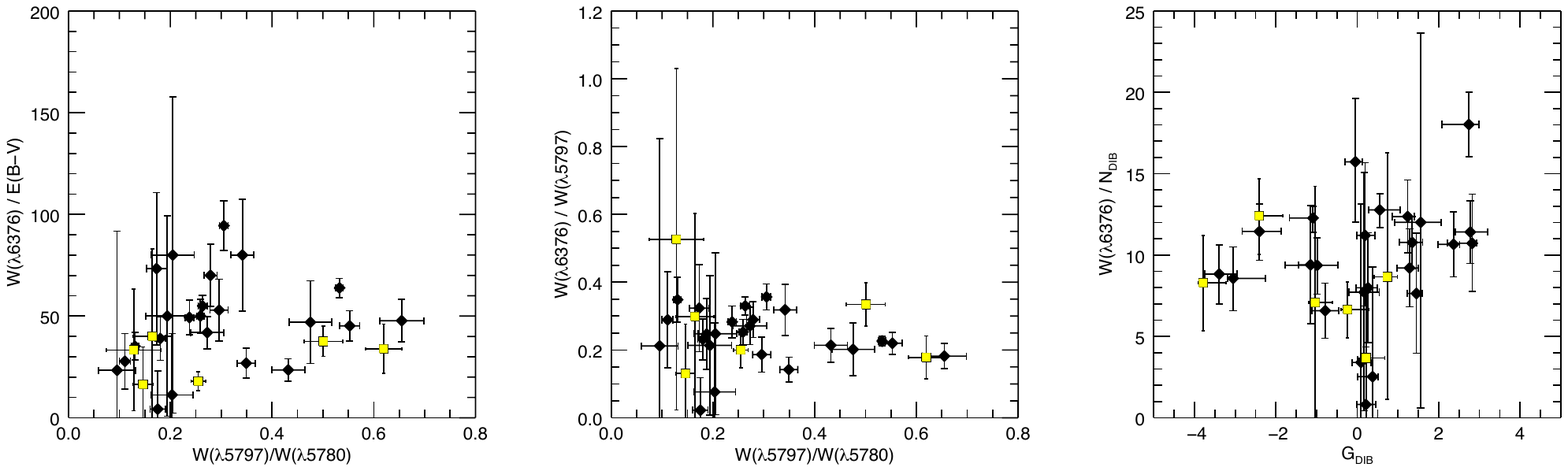}\\
           \includegraphics[width=16cm, trim={2cm, 10cm, 1.5cm, 11cm}]{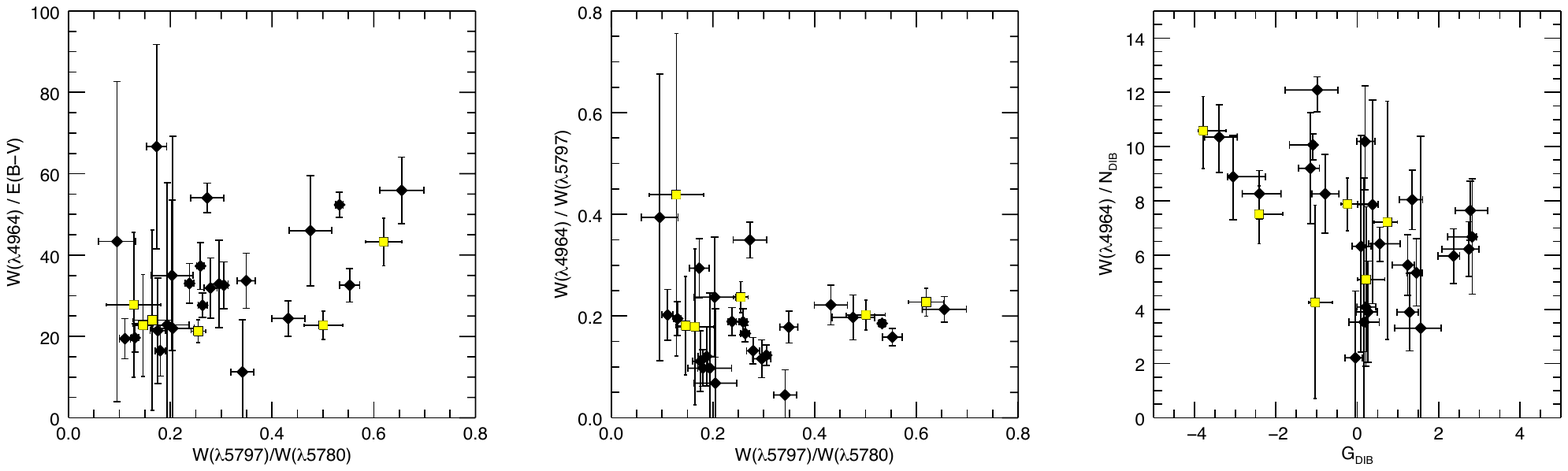}\\
           \includegraphics[width=16cm, trim={2cm, 10cm, 1.5cm, 11cm}]{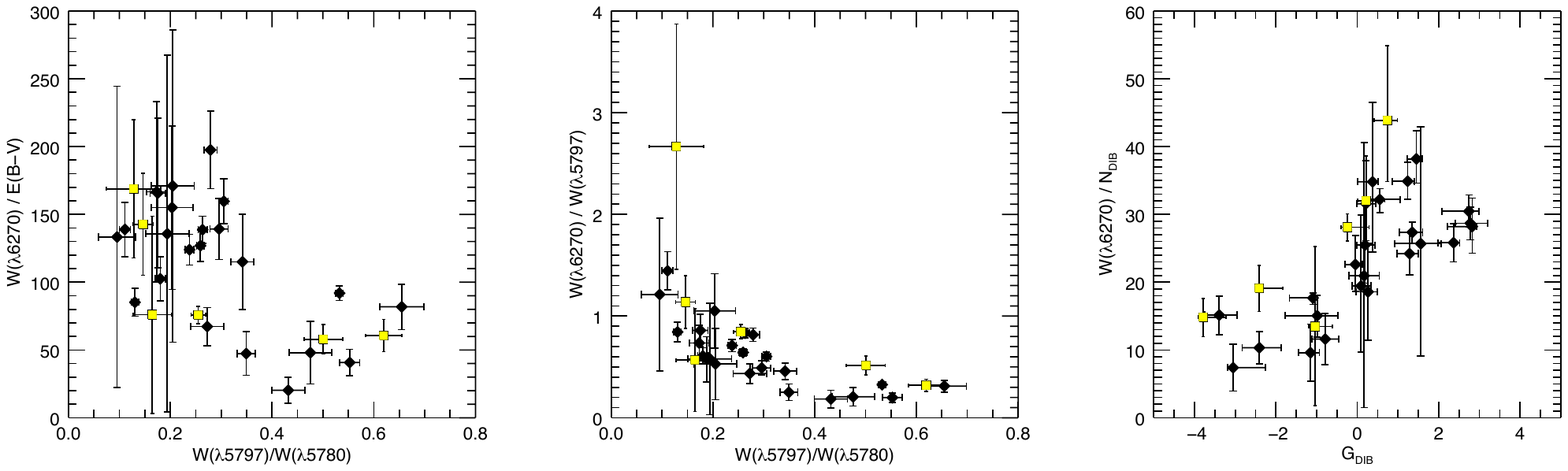}
\end{array}$    
\end{center}
\caption{\label{fig:enviro} The DIBs (top to bottom) $\lambda$6376,
  $\lambda$4964, and $\lambda$6270  (\textit{Left}:) normalized to
  E(B-V) as a function of W($\lambda$5797)/W($\lambda$5780);
  (\textit{Centre}:)  normalized to W($\lambda$5797) as a function of
  W($\lambda$5797)/W($\lambda$5780); (\textit{Right}:) normalized to
  N$_{\text{DIB}}$ as a function of G$_{\text{DIB}}$. Be stars are
  indicated by yellow squares. HD~143275 has been excluded from the
  left plots because it has an E(B-V) value of zero.} 
\end{figure*}

The key result obtained through this exercise is that a variety of
DIBs in single cloud lines of sight representing very different
environments can be described remarkably well by just four parameters
-- a surprising fact considering the large number of DIBs and the lack
of strong correlations among them. This implies that differences in
interstellar environments are not driven by the DIBs themselves, but
rather, the DIBs react to a limited number of parameters that dictate
ISM conditions. Each DIB has a unique response to these conditions,
and thus we see a unique DIB spectrum in all lines of sight. Note that
also \citet{Lan:SDSS_DIBs} could reduce much of the variation in the
average DIB properties to just two parameters.

DIB strength is predominantly determined by the parameter
N$_{\text{DIB}}$. In Figure~\ref{fig:dibs_vs_ndib}, we show the
equivalent width of each DIB included in our sample as a function of
N$_{\text{DIB}}$. From this figure, we can gain two important
insights. The first is that we can see how capable N$_{\text{DIB}}$ is
of predicting the strength of each DIB. Those DIBs that correlate very
strongly (e.g., $\lambda$5797, $\lambda$5850, $\lambda$5494) can be
predicted by a single parameter -- that is to say, N$_{\text{DIB}}$ is
singularly capable of describing their observed strengths. For DIBs
that correlate less strongly (e.g., $\lambda$5769, $\lambda$6284,
$\lambda$5513), there are other parameters (such as G$_{\text{DIB}}$
and GTD) playing a significant role. The second insight gained from
Figure~\ref{fig:dibs_vs_ndib} is that we can infer the order in which
DIBs form. For example, if the line of best fit intersects the x-axis
at a small value (e.g., $\lambda$6284, $\lambda$4428, $\lambda$5780,
$\lambda$6270) then these carriers start to form when
N$_{\text{DIB}}$, the collective level of DIB-producing material, is
still small. In other words, these carriers form before the others. On
the other hand, DIBs like $\lambda$5850, $\lambda$4964, and
$\lambda$5513 start forming once the aforementioned DIB carriers are
already present in appreciable amounts. These differences might
reflect the molecular complexity of the carriers (i.e., simple species
form at lower N$_{\text{DIB}}$ values and may act as precursors for
those species forming at higher N$_{\text{DIB}}$ values) or ionization
potentials. It should be noted that negative N$_{\text{DIB}}$
intercepts are permissible, since N$_{\text{DIB}}$ acts like the
column density of the ``average" DIB carrier.

The second most important parameter for determining DIB strength is
the amount of UV radiation. To specifically investigate how DIBs
respond to changes in UV exposure, we must first normalize DIB
strengths to the amount of material. We illustrate this in three
different ways in Figure~\ref{fig:enviro} for three DIBs:
$\lambda$6376 (whose variation is primarily determined by
N$_{\text{DIB}}$), $\lambda$4964 (a C$_2$-DIB), and $\lambda$6270
(which has quite a bit of variability unaccounted for by
N$_{\text{DIB}}$). Similar plots for our full set of DIBs are shown in
Fig.~\ref{fig:DIB_enviro} in the appendix. In the leftmost plots, we
employ the common practice of normalizing DIBs to E(B-V), and plot
W($\lambda$5797)/W($\lambda$5780) (the variable that best correlates
with G$_{\text{DIB}}$) on the x-axis. In the center plots, we instead
normalize DIBs to W($\lambda$5797) -- the variable which most strongly
correlates with N$_{\text{DIB}}$. Finally, in the rightmost plots,
DIBs are normalized to N$_{\text{DIB}}$ directly and are plotted as a
function of G$_{\text{DIB}}$.

A comparison of the left and center plots reveals the consequences of
normalizing DIBs to E(B-V) versus W($\lambda$5797). The latter situation
tends to yield clearer trends compared to the former. This is most
apparent for $\lambda$6270, which exhibits clear environmental
behavior. However, even when the main trend is less discernible (e.g.,
for $\lambda$6376), a reduced amount of scatter is still noted.

The plots on the right provide a clearer picture of DIB environmental
behavior, although they have the disadvantage of not being
reproducible without individually measuring the variables
corresponding to the 23 terms in Equation \ref{eqn:PC2_eqn} or else
performing PCA. Note that the x-axis is reversed in these plots,
compared to the left and center ones (i.e., PC$_1$ increases for
decreasing W($\lambda$5797)/W($\lambda$5780) values). $\lambda$6270
has a positive slope, suggesting that it is an ionized or
dehydrogenated species. As G$_{\text{DIB}}$ increases, more carriers
become ionized dehydrogenated and this DIB becomes stronger. It makes
sense that $\lambda$6270 would be strongly influenced by
G$_{\text{DIB}}$ since, referring to Figure~\ref{fig:dibs_vs_ndib},
N$_{\text{DIB}}$ is unable to account for a sizeable portion of this
DIB's variation. The remaining variation must be driven by the
subsequent PCs and hence, G$_{\text{DIB}}$ shows a strong trend for
$\lambda$6270. $\lambda$4964, on the other hand, displays the opposite
behavior: As G$_{\text{DIB}}$ increases, the strength of $\lambda$4964
decreases, suggesting that the carrier prefers more sheltered
environments (e.g., lower G$_0$, higher densities, and cooler
temperatures). For $\lambda$6376, the trend with G$_{\text{DIB}}$ is
not particularly strong, suggesting that this feature does not
systematically vary with G$_0$ over the range of UV exposures covered
in this study. Incidentally, this DIB has a very small projection on
PC$_2$ (see Figure~\ref{fig:PC1-PC2}).

The physical interpretation for PC$_3$ is less certain. Because of
this fact, we are limited in our ability to infer characteristics of
the DIB carriers based on their behavior with respect to PC$_3$. It
seems clear though that PC$_3$ is tracing a property of the dust or
its relation with the gas (such as e.g. the gas-to-dust ratio),
suggesting that dust may play a role in the formation and/or
excitation of some of the DIB carriers.
	
\section{Conclusions}
\label{sec:Conclusions}

We performed a principal component analysis on a selection of DIBs and
line of sight parameters measured for single-cloud sightlines. We
found that the majority of DIB variations can be attributed to four
parameters. The variable that most strongly determines DIB strength is
the amount of DIB-producing material in the line of sight, a parameter
that is traced extremely well by W($\lambda$5797). The second most
important parameter is the level of UV exposure, which is reasonably
well-approximated by W($\lambda$5797)/W($\lambda$5780) in $\zeta$
sightlines. W($\lambda$5797) is unaffected by this parameter, and
therefore changes in W($\lambda$5797)/W($\lambda$5780) primarily reflect
changes in the $\lambda$5780 carrier. The third is presumably related
to the dust properties in the line of sight, with the gas-to-dust
ratio being one possibility. Finally, the fourth parameter is related
to the depletions in the line of sight, although we do not offer a
physical interpretation beyond this.
	
The work presented in this paper is only a first attempt at
discriminating the different parameters that drive the variations in
the DIBs and line of sight parameters. Our results show that there is
great potential in this method to recognize the processes at play, and
to turn the DIBs themselves into powerful probes of the conditions in
their surroundings. With the current measurement errors however, only
the first three principal components can be somewhat confidently
identified; the uncertainties on the remaining components are too
large for a meaningful interpretation. At this point, it is not clear
whether the uncerainties on these components is the result of
intrinsic differences in the DIBs (e.g. different carrier abundances),
or whether it is the direct consequence of measurement errors. In
future work, this will become clear when using a much larger sample of
DIBs from high signal-to-noise observations of many more sightlines in
the context of the European Southern Observatory Diffuse Interstellar
Bands Large Exploration Survey (EDIBLES). With a larger number of
observations, one could also consider a different PCA that can then
result in ``principal eigenspectra''. While such an exercise would be
very interesting, it can only be expected to work if the different
spectral sources (stellar, interstellar and telluric) can be reliably
separated.

\acknowledgments We thank the anonymous referee for valuable comments
that improved the clarity of this manuscript. TE, JC, NHB, and AS
acknowledge the support from the Natural Sciences and Engineering
Research Council of Canada (NSERC) through a Discovery grant. This
research has made use of NASA's Astrophysics Data System and of the
SIMBAD database, operated at CDS, Strasbourg, France. 

\software{IDL; molecfit \citep{Molecfit1,Molecfit2};
  MPFITEXY \citep{MPFITEXY}; MPFIT package \citep{MPFIT}}

\facility{VLT:Kueyen (UVES), OHP:1.93m (ELODIE)}

\newpage
\appendix
\section{Measurements}
Below is an overview of all the DIB equivalent width measurements that
are used in this paper. Note that the uncertainties on the
measurements were obtained by using a Monte Carlo approach to simulate
positioning the continuum. Table~\ref{table:EW_meas} lists the
measurements for the $\lambda\lambda$4428, 4964, 5494, 5513, 5545,
5546, 5769 and 5780 DIBs; Table~\ref{table:EW_meas2} lists the same
for the $\lambda\lambda$5797, 5850, 6196, 6270, 6284, 6376, 6379 and
6614 DIBs.

\begin{deluxetable}{lcrrrrrrr}
\tablecaption{\label{table:EW_meas} Equivalent widths and
    uncertainties of the $\lambda\lambda$4428, 4964, 5494, 5513, 5545, 5546, 5769 and 5780 DIBs.}
\tablehead{
\colhead{Target} &
\colhead{$\lambda$4428}           & \colhead{$\lambda$4964}      &
\colhead{$\lambda$5494}          & \colhead{$\lambda$5513}  &
\colhead{$\lambda$5545}          & \colhead{$\lambda$5546}    &
\colhead{$\lambda$5769}  & \colhead{$\lambda$5780}
}
\startdata
HD 15137 &1163 $\pm$ $^{106}_{115}$ &7.9 $\pm$ 2.5 &11.1 $\pm$ 2.2 &2.1 $\pm$ 3.0 &6.9 $\pm$ 1.9 &0.0 $\pm$ 1.9 &3.9 $\pm$ 1.7 &230.1 $\pm$ 9.1\\
HD 22951 &471 $\pm$ $^{68}_{76}$ &6.4 $\pm$  1.1 &2.0 $\pm$ 1.1 &1.3 $\pm$ 1.5 &6.2 $\pm$ 0.7 &3.6 $\pm$ 1.0 &0.7 $\pm$ 0.8 &102.8 $\pm$ 3.6\\
HD 23180 &403 $\pm$ $^{45}_{47}$ &12.3 $\pm$ 1.4 &6.4 $\pm$ 0.2 &10.7 $\pm$ 1.7 &10.3 $\pm$ 1.5 &5.4 $\pm$ 1.5 &7.2 $\pm$ 1.3 &88.1 $\pm$ 5.0\\
HD 23630 &325 $\pm$ $^{48}_{39}$ &1.2 $\pm$ 1.0 &2.4 $\pm$ 0.9 &0.2 $\pm$ 1.5 &0.8 $\pm$ 1.1 &1.5 $\pm$ 1.0 &2.3 $\pm$ 0.9 &40.7 $\pm$ 4.8\\
HD 24398 &450 $\pm$ $^{61}_{70}$ &8.8 $\pm$ 0.9 &5.4 $\pm$ 1.0 &5.8 $\pm$ 1.1 &6.1 $\pm$ 0.6 &3.3 $\pm$ 1.0 &2.5 $\pm$ 0.7 &100.4 $\pm$ 2.7\\
HD 24534 &402 $\pm$ $^{49}_{55}$ &13.4 $\pm$ 1.6 &7.6 $\pm$ 1.2 &5.3 $\pm$ 1.9 &9.4 $\pm$ 1.2 &4.8 $\pm$ 1.6 &7.1 $\pm$ 1.1 &95.1 $\pm$ 5.0\\
HD 24760 &322 $\pm$ $^{41}_{30}$ &1.5 $\pm$ 0.8 &3.3 $\pm$ 0.8 &1.1 $\pm$ 1.0 &1.5 $\pm$ 0.9 &0.2 $\pm$ 0.8 &1.6 $\pm$ 0.6 &77.0 $\pm$ 3.4\\
HD 24912 &949 $\pm$ $^{89}_{65}$ &9.7 $\pm$ 1.3 &7.0 $\pm$ 1.0 &2.7 $\pm$ 1.2 &8.9 $\pm$ 1.0 &2.4 $\pm$ 1.2 &2.4 $\pm$ 0.8 &198.3 $\pm$ 3.1\\
HD 27778 &490 $\pm$ $^{74}_{58}$ &8.3 $\pm$ 1.4 &4.6 $\pm$ 1.6 &3.6 $\pm$ 1.4 &8.0 $\pm$ 1.1 &4.5 $\pm$ 1.0 &2.2 $\pm$ 1.0 &86.6 $\pm$ 4.6\\
HD 35149 &254  $\pm$ $^{43}_{38}$ &2.8 $\pm$ 1.3 &2.8 $\pm$ 1.7 &1.0 $\pm$ 1.9 &2.6 $\pm$ 1.3 &0.0 $\pm$ 1.4 &1.7 $\pm$ 1.5 &58.0 $\pm$ 5.5\\
HD 35715 &221 $\pm$ $^{47}_{23}$ &1.3 $\pm$ 0.8 &1.1 $\pm$ 0.8 &1.2 $\pm$ 0.9 &1.1 $\pm$ 0.8 &0.7 $\pm$ 0.9 &0.7 $\pm$ 0.7 &34.6 $\pm$ 3.6\\
HD 36822 &483  $\pm$ $^{78}_{69}$ &1.6 $\pm$ 2.4 &1.4 $\pm$ 2.8 &2.9 $\pm$ 3.0 &2.0 $\pm$ 2.4 &2.9 $\pm$ 2.4 &1.0 $\pm$ 2.0 &84.5 $\pm$ 9.6\\
HD 36861 &402  $\pm$ $^{49}_{86}$ &4.6 $\pm$ 1.0 &3.2 $\pm$ 1.0 &4.4 $\pm$ 1.1 &3.2 $\pm$ 0.9 &3.2 $\pm$ 0.9 &1.5 $\pm$ 0.7 &49.0 $\pm$ 3.5\\
HD 40111 &739  $\pm$ $^{109}_{81}$ &2.2 $\pm$ 4.7 &2.7 $\pm$ 4.9 &0.0 $\pm$ 7.2 &3.6 $\pm$ 4.4 &3.2 $\pm$ 4.9 &3.3 $\pm$ 3.7 &157.7 $\pm$ 19.5\\
HD 110432 &880  $\pm$ $^{64}_{45}$ &8.3 $\pm$ 1.0 &4.1 $\pm$ 1.0 &3.8 $\pm$ 1.4 &5.2 $\pm$ 1.0 &1.8 $\pm$ 0.8 &0.3 $\pm$ 0.8 &137.3 $\pm$ 3.7\\
HD 143275 &383  $\pm$ $^{21}_{12}$ &2.1 $\pm$ 1.0 &5.1 $\pm$ 0.1 &2.1 $\pm$ 1.5 &5.2 $\pm$ 1.1 &1.4 $\pm$ 1.2 &1.9 $\pm$ 1.1 &92.7 $\pm$ 4.2\\
HD 144217 &430  $\pm$ $^{54}_{38}$ &3.5 $\pm$ 0.8 &2.6 $\pm$ 1.0 &1.1 $\pm$ 1.6 &4.1 $\pm$ 1.1 &1.0 $\pm$ 1.0 &0.7 $\pm$ 1.1 &156.0 $\pm$ 4.9\\
HD 145502 &583  $\pm$ $^{50}_{48}$ &3.3 $\pm$ 1.2 &6.3 $\pm$ 2.0 &2.8 $\pm$ 2.5 &4.4 $\pm$ 1.0 &2.0 $\pm$ 1.2 &3.0 $\pm$ 0.9 &186.9 $\pm$ 5.2\\
HD 147165 &872  $\pm$ $^{50}_{53}$ &6.1 $\pm$ 1.0 &8.2 $\pm$ 1.5 &5.1 $\pm$ 1.6 &4.5 $\pm$ 1.0 &1.9 $\pm$ 1.2 &0.8 $\pm$ 1.1 &240.0 $\pm$ 4.2\\
HD 147933 &1254  $\pm$ $^{121}_{77}$ &20.0 $\pm$ 0.8 &7.6 $\pm$ 0.5 &13.8 $\pm$ 0.7 &8.3 $\pm$ 0.5 &6.9 $\pm$ 0.6 &11.7 $\pm$ 2.8 &209.8 $\pm$ 16.1\\
HD 149757 &576  $\pm$ $^{52}_{47}$ &6.6 $\pm$ 0.9 &5.3 $\pm$ 1.1 &3.0 $\pm$ 1.3 &5.7 $\pm$ 0.9 &2.5 $\pm$ 0.8 &2.8 $\pm$ 1.1 &65.1 $\pm$ 3.8\\
HD 164284 &686  $\pm$ $^{73}_{53}$ &2.5 $\pm$ 1.3 &1.8 $\pm$ 1.4 &2.3 $\pm$ 1.8 &3.4 $\pm$ 1.1 &1.5 $\pm$ 1.4 &0.7 $\pm$ 1.0 &94.4 $\pm$ 4.4\\
HD 170740 &834  $\pm$ $^{107}_{91}$ &10.5 $\pm$ 1.0 &10.6 $\pm$ 1.0 &8.6 $\pm$ 1.5 &11.3 $\pm$ 1.0 &4.6 $\pm$ 1.0 &2.4 $\pm$ 0.8 &240.3 $\pm$ 4.0\\
HD 198478 &1592  $\pm$ $^{191}_{108}$ &14.2 $\pm$ 2.0 &10.5 $\pm$ 1.8 &5.8 $\pm$ 2.1 &11.8 $\pm$ 1.4 &5.0 $\pm$ 1.5 &1.5 $\pm$ 1.3 &315.6 $\pm$ 5.8\\
HD 202904 &541 $\pm$ $^{92}_{67}$ &2.5 $\pm$ 1.5 &2.6 $\pm$ 1.5 &1.8 $\pm$ 1.6 &1.0 $\pm$ 1.0 &1.1 $\pm$ 1.2 &1.0 $\pm$ 1.1 &44.5 $\pm$ 4.6\\
HD 207198 &1282  $\pm$ $^{67}_{89}$ &24.6 $\pm$ 1.0 &19.3 $\pm$ 0.9 &16.6 $\pm$ 1.1 &20.5 $\pm$ 0.9 &9.9 $\pm$ 0.9 &9.8 $\pm$ 0.7 &249.0 $\pm$ 2.8\\
HD 209975 &1032  $\pm$ $^{182}_{74}$ &8.8 $\pm$ 1.4 &13.0 $\pm$ 1.5 &1.9 $\pm$ 1.9 &11.2 $\pm$ 0.7 &4.6 $\pm$ 1.6 &0.4 $\pm$ 1.3 &234.2 $\pm$ 4.7\\
HD 214680 &361  $\pm$ $^{55}_{64}$ &0.9 $\pm$ 1.0 &4.4 $\pm$ 0.8 &1.8 $\pm$ 1.4 &1.7 $\pm$ 1.0 &2.0 $\pm$ 0.9 &0.6 $\pm$ 0.5 &58.8 $\pm$ 2.8\\
HD 214993 &232  $\pm$ $^{63}_{47}$ &4.0 $\pm$ 0.7 &0.2 $\pm$ 1.2 &1.6 $\pm$ 1.3 &1.4 $\pm$ 1.0 &1.2 $\pm$ 0.9 &0.6 $\pm$ 0.8 &78.6 $\pm$ 4.8\\
HD 218376 &766  $\pm$ $^{64}_{108}$ &5.1 $\pm$ 1.0 &5.9 $\pm$ 1.1 &3.9 $\pm$ 1.2 &6.2 $\pm$ 0.8 &1.1 $\pm$ 1.1 &1.1 $\pm$ 0.8 &138.7 $\pm$ 4.4\\
\enddata
\tablecomments{All measurements in m\AA.}
\end{deluxetable}

\begin{deluxetable}{crrrrrrrr}
\tablecaption{\label{table:EW_meas2}Equivalent widths and
    uncertainties of the $\lambda\lambda$5797, 5850, 6196, 6270,
  6284, 6376, 6379 and 6614 DIBs. }
\tablehead{
\colhead{Target} &
\colhead{$\lambda$5797} & \colhead{$\lambda$5850} &
\colhead{$\lambda$6196} & \colhead{$\lambda$6270} &
\colhead{$\lambda$6284} & \colhead{$\lambda$6376} &
\colhead{$\lambda$6379} & \colhead{$\lambda$6614} 
}
\startdata
HD 15137 &68.1 $\pm$ 3.1 &20.8 $\pm$ 2.8 &19.9 $\pm$ 2.7 &33.4 $\pm$ 4.6 &298.6 $\pm$ 19.4 &12.7 $\pm$ 3.4 &36.2 $\pm$ 4.2 &80.6 $\pm$ 4.1\\
HD 22951 &35.9 $\pm$ 1.3 &18.8 $\pm$ 1.2 &10.5 $\pm$ 2.2 &9.0 $\pm$ 2.9 &130.8 $\pm$ 8.5 &5.1 $\pm$ 1.3 &23.8 $\pm$ 1.5 &41.0 $\pm$ 2.1\\
HD 23180 &57.7 $\pm$ 2.0 &27.8 $\pm$ 1.3 &12.8 $\pm$ 1.9 &18.0 $\pm$ 3.3 &95.4 $\pm$ 9.4 &10.5 $\pm$ 2.1 &41.3 $\pm$ 3.0 &53.7 $\pm$ 3.4\\
HD 23630 &6.7 $\pm$ 1.3 &1.5 $\pm$ 1.0 &1.9 $\pm$ 1.3 &3.8 $\pm$ 3.3 &21.0 $\pm$ 7.7 &2.0 $\pm$ 2.0 &3.0 $\pm$ 2.1 &8.9 $\pm$ 2.8\\
HD 24398 &55.5 $\pm$ 1.3 &27.3 $\pm$ 1.1 &15.2 $\pm$ 1.2 &11.0 $\pm$ 2.5 &94.1 $\pm$ 6.7 &12.2 $\pm$ 1.8 &46.3 $\pm$ 2.5 &59.3 $\pm$ 1.9\\
HD 24534 &58.9 $\pm$ 1.3 &29.0 $\pm$ 1.4 &15.2 $\pm$ 1.4 &18.8 $\pm$ 3.5 &78.2 $\pm$ 8.2 &10.5 $\pm$ 3.7 &40.3 $\pm$ 2.3 &66.1 $\pm$ 2.4\\
HD 24760 &13.5 $\pm$ 1.0 &2.9 $\pm$ 0.8 &6.0 $\pm$ 1.2 &11.6 $\pm$ 2.0 &105.9 $\pm$ 5.5 &0.3 $\pm$ 1.3 &8.2 $\pm$ 1.5 &23.3 $\pm$ 2.1\\
HD 24912 &51.4 $\pm$ 1.2 &22.3 $\pm$ 1.7 &21.7 $\pm$ 1.0 &33.0 $\pm$ 1.7 &272.4 $\pm$ 9.6 &13.0 $\pm$ 1.9 &30.1 $\pm$ 2.3 &79.7 $\pm$ 1.8\\
HD 27778 &37.4 $\pm$ 2.0 &12.7 $\pm$ 1.3 &10.8 $\pm$ 1.5 &6.9 $\pm$ 3.2 &117.8 $\pm$ 10.2 &8.0 $\pm$ 1.8 &17.4 $\pm$ 2.1 &45.7 $\pm$ 2.7\\
HD 35149 &11.8 $\pm$ 2.1 &6.8 $\pm$ 1.3 &7.1 $\pm$ 1.9 &12.4 $\pm$ 3.7 &78.0 $\pm$ 14.4 &0.9 $\pm$ 2.4 &6.0 $\pm$ 3.3 &21.9 $\pm$ 4.6\\
HD 35715 &3.3 $\pm$ 1.2 &0.5 $\pm$ 0.7 &2.4 $\pm$ 1.1 &4.0 $\pm$ 2.0 &55.4 $\pm$ 8.4 &0.7 $\pm$ 2.0 &2.8 $\pm$ 1.9 &9.5 $\pm$ 1.9\\
HD 36822 &16.4 $\pm$ 3.1 &3.7 $\pm$ 2.2 &8.1 $\pm$ 3.1 &9.5 $\pm$ 8.8 &106.6 $\pm$ 15.9 &3.5 $\pm$ 3.3 &10.1 $\pm$ 5.4 &18.0 $\pm$ 6.2\\
HD 36861 &23.3 $\pm$ 1.2 &12.3 $\pm$ 0.8 &4.9 $\pm$ 1.0 &4.8 $\pm$ 2.1 &51.6 $\pm$ 10.8 &4.7 $\pm$ 1.8 &6.2 $\pm$ 1.4 &14.9 $\pm$ 1.8\\
HD 40111 &32.3 $\pm$ 5.3 &3.6 $\pm$ 3.1 &13.0 $\pm$ 5.6 &17.1 $\pm$ 1..0 &211.1 $\pm$ 22.5 &8.0 $\pm$ 7.6 &12.9 $\pm$ 9.5 &41.1 $\pm$ 9.6\\
HD 110432 &35.0 $\pm$ 1.7 &19.4 $\pm$ 1.0 &18.0 $\pm$ 1.0 &29.6 $\pm$ 2.0 &185.1 $\pm$ 5.1 &7.0 $\pm$ 1.8 &32.4 $\pm$ 1.8 &74.3 $\pm$ 2.1\\
HD 143275 &17.4 $\pm$ 1.3 &6.3 $\pm$ 1.1 &7.6 $\pm$ 0.9 &10.0 $\pm$ 3.8 &118.9 $\pm$ 13.1 &4.3 $\pm$ 1.8 &10.1 $\pm$ 3.0 &23.9 $\pm$ 1.6\\
HD 144217 &17.3 $\pm$ 1.6 &6.5 $\pm$ 1.1 &13.5 $\pm$ 1.5 &25.0 $\pm$ 2.3 &159.3 $\pm$ 9.1 &5.0 $\pm$ 2.4 &14.0 $\pm$ 3.6 &50.9 $\pm$ 1.7\\
HD 145502 &33.7 $\pm$ 1.7 &12.2 $\pm$ 1.2 &14.1 $\pm$ 2.6 &20.5 $\pm$ 2.5 &199.6 $\pm$ 8.8 &7.8 $\pm$ 2.0 &30.0 $\pm$ 2.0 &58.8 $\pm$ 2.5\\
HD 147165 &31.3 $\pm$ 1.6 &16.7 $\pm$ 1.1 &17.5 $\pm$ 1.1 &26.4 $\pm$ 2.7 &214.2 $\pm$ 7.7 &10.9 $\pm$ 2.0 &21.1 $\pm$ 2.0 &61.3 $\pm$ 2.3\\
HD 147933 &57.2 $\pm$ 5.3 &30.6 $\pm$ 2.6 &17.0 $\pm$ 2.7 &24.9 $\pm$ 5.0 &173.8 $\pm$ 16.9 &15.5 $\pm$ 2.8 &28.0 $\pm$ 3.7 &62.5 $\pm$ 3.6\\
HD 149757 &32.6 $\pm$ 1.6 &14.2 $\pm$ 1.1 &10.3 $\pm$ 1.2 &16.8 $\pm$ 2.9 &72.0 $\pm$ 6.9 &10.9 $\pm$ 2.0 &16.7 $\pm$ 1.9 &46.4 $\pm$ 2.0\\
HD 164284 &13.8 $\pm$ 1.7 &0.4 $\pm$ 1.3 &6.8 $\pm$ 1.5 &15.7 $\pm$ 3.0 &111.3 $\pm$ 9.2 &1.8 $\pm$ 2.0 &11.3 $\pm$ 2.2 &26.9 $\pm$ 2.7\\
HD 170740 &63.3 $\pm$ 1.8 &24.6 $\pm$ 1.1 &26.3 $\pm$ 1.2 &52.7 $\pm$ 2.6 &249.6 $\pm$ 9.9 &20.9 $\pm$ 1.6 &60.7 $\pm$ 1.7 &122.4 $\pm$ 2.2\\
HD 198478 &75.0 $\pm$ 2.2 &34.6 $\pm$ 1.6 &33.1 $\pm$ 1.5 &53.3 $\pm$ 4.2 &379.5 $\pm$ 11.6 &21.2 $\pm$ 3.5 &46.7 $\pm$ 4.1 &130.6 $\pm$ 3.4\\
HD 202904 &5.7 $\pm$ 2.3 &1.9 $\pm$ 1.7 &3.6 $\pm$ 1.8 &15.2 $\pm$ 3.1 &82.2 $\pm$ 10.6 &3.0 $\pm$ 2.6 &11.7 $\pm$ 3.5 &18.4 $\pm$ 2.7\\
HD 207198 &132.6 $\pm$ 1.1 &61.1 $\pm$ 0.7 &32.3 $\pm$ 1.0 &43.2 $\pm$ 1.7 &227.2 $\pm$ 9.6 &30.0 $\pm$ 1.8 &71.8 $\pm$ 2.1 &121.8 $\pm$ 1.9\\
HD 209975 &71.5 $\pm$ 1.4 &26.5 $\pm$ 1.6 &26.9 $\pm$ 4.5 &43.1 $\pm$ 3.1 &240.2 $\pm$ 10.0 &25.5 $\pm$ 2.7 &45.5 $\pm$ 2.6 &114.1 $\pm$ 3.1\\
HD 214680 &20.1 $\pm$ 0.9 &3.9 $\pm$ 0.9 &5.4 $\pm$ 1.0 &9.2 $\pm$ 1.6 &68.7 $\pm$ 7.9 &6.4 $\pm$ 1.5 &4.5 $\pm$ 1.4 &16.1 $\pm$ 2.0\\
HD 214993 &13.6 $\pm$ 1.3 &0.9 $\pm$ 0.7 &7.6 $\pm$ 1.4 &10.0 $\pm$ 2.2 &107.1 $\pm$ 10.0 &4.4 $\pm$ 1.7 &13.9 $\pm$ 1.7 &18.0 $\pm$ 2.3\\
HD 218376 &38.7 $\pm$ 1.3 &17.2 $\pm$ 1.0 &14.2 $\pm$ 1.2 &31.6 $\pm$ 2.3 &175.7 $\pm$ 10.0 &11.2 $\pm$ 2.0 &37.0 $\pm$ 2.2 &66.0 $\pm$ 2.2\\
\enddata
\tablecomments{All measurements in m\AA.}
\end{deluxetable}

\section{The main PCs}
\setcounter{equation}{0}
\renewcommand{\theequation}{A\arabic{equation}}

As is discussed in the main body of the paper, only the first four
principal components can be interpreted in a meaningful way. While the
method and the general formalism is presented in Sect.~\ref{sec:PCA},
it is useful and insightful to have the full expressions (i.e. the
coefficients $a$ in Eq.~(\ref{Eq:PC_vector})) for the principal
components in terms of the original variables as well as in terms of
the standardized variables. Below are the equations for the leading
four PCs in terms of the standardized variables ($z_1$, $z_2$, ...,
$z_{23}$) and then re-expressed in terms of the original, measurable
variables. See Sect.~\ref{sec:PCA} for details. 

\begin{align}
\label{eqn:PC1_eqn}
PC_{1} =& 0.208(z_1) + 0.239 (z_2) + 0.230 (z_3) + 0.203 (z_4) + 0.243 (z_5) + 0.206 (z_6) + 0.157(z_7)\\
&+ 0.207 (z_8) + 0.245 (z_9) + 0.244 (z_{10}) + 0.238 (z_{11}) + 0.208 (z_{12})+ 0.177 (z_{13})  0.240 (z_{14})\nonumber \\
&+ 0.234 (z_{15})  + 0.236 (z_{16}) + 0.232 (z_{17}) + 0.186 (z_{18}) + 0.203 (z_{19}) + 0.232 (z_{20}) + 0.110 (z_{21})\nonumber \\
&+ 0.096 (z_{22}) + 0.118 (z_{23}) \nonumber \\
=& (5.92\times 10^{-4})W(\lambda4428) + (4.18\times 10^{-2})W(\lambda4964) + (5.56\times 10^{-2})W(\lambda 5494) + (5.24 \times 10^{-2}) W(\lambda 5513) \nonumber \\
&+ (5.63 \times 10^{-2}) W(\lambda 5545) + (9.36 \times 10^{-2}) W(\lambda5546) + (5.62 \times 10^{-2}) W(\lambda 5769) + (2.69 \times 10^{-3}) W(\lambda 5780)\nonumber \\
&+ (8.83 \times 10^{-3}) W(\lambda 5797) + (1.80 \times 10^{-2}) W(\lambda5850) + (2.85 \times 10^{-2}) W(\lambda 6196) + (1.49 \times 10^{-2}) W(\lambda6270)\nonumber\\
&+ (2.10 \times 10^{-3})W(\lambda 6284) + (3.24 \times 10^{-2}) W(\lambda6376) + (1.30 \times 10^{-2}) W(\lambda 6379) + (6.70 \times 10^{-3})W(\lambda6614)\nonumber\\
&+ (1.75)E(B-V) + (2.03 \times 10^{-22}) N(H \textsc{i}) + (7.80 \times 10 ^{-22}) N(H_2) + (1.95 \times 10^{-22}) N(H)\nonumber\\
&+ (4.72 \times 10^{-1}) f(H_2) + (4.49 \times 10^{-1}) F_{\star} + (7.50 \times 10^{-1}) W(\lambda5797)/W(\lambda5780) - 6.62 \nonumber
\end{align}

\begin{align}
\label{eqn:PC2_eqn}
 PC_2 =& 0.245(z_1) - 0.137(z_2) + 0.085(z_3) -0.185(z_4) -0.058(z_5)  -0.203 (z_6) -0.238(z_7)\\
 &+ 0.306 (z_8) -0.020(z_9) -0.085(z_{10}) + 0.168(z_{11})11 + 0.264(z_{12}) + 0.361(z_{13}) + 0.077(z_{14})\nonumber \\
 &-0.008 (z_{15}) + 0.145 (z_{16}) -0.055(z_{17}) + 0.109(z_{18})  -0.226 (z_{19})  -0.015 (z_{20}) -0.334(z_{21})\nonumber \\
 &-0.287(z_{22})  -0.403(z_{23}) \nonumber \\
=&(6.98\times 10^{-4})W(\lambda4428) - (2.40\times 10^{-2})W(\lambda4964) + (2.06 \times 10^{-2})W(\lambda 5494) - (4.77 \times 10^{-2}) W(\lambda 5513) \nonumber \\
&- (1.35 \times 10^{-2}) W(\lambda 5545) + (9.21 \times 10^{-2}) W(\lambda5546) - (8.53 \times 10^{-2}) W(\lambda 5769) + (3.97 \times 10^{-3}) W(\lambda 5780)\nonumber \\
&- (7.27 \times 10^{-4}) W(\lambda 5797) - (6.23 \times 10^{-3}) W(\lambda5850) + (2.02 \times 10^{-2}) W(\lambda 6196) + (1.89 \times 10^{-2}) W(\lambda6270)\nonumber\\
&+ (4.29 \times 10^{-3})W(\lambda 6284) + (1.04 \times 10^{-2}) W(\lambda6376) - (4.29 \times 10^{-4}) W(\lambda 6379) + (4.11 \times 10^{-3})W(\lambda6614)\nonumber\\
&- (4.13 \times 10^{-1})E(B-V) + (1.19 \times 10^{-22}) N(H \textsc{i}) - (8.69 \times 10 ^{-22}) N(H_2) - (1.23 \times 10^{-23}) N(H)\nonumber\\
&- (1.43) f(H_2) - (1.34) F_{\star} - (2.57) W(\lambda5797)/W(\lambda5780) - 0.706 \nonumber
\end{align}

\begin{align}
\label{eqn:PC3_eqn}
PC_3 =& 0.033(z_1) + 0.138(z_2) + 0.036(z_3) + 0.287(z_4)  -0.024(z_5) + 0.178(z_6) + 0.396(z_7)\\
& -0.022(z_8) + 0.008(z_9) + 0.047(z_{10})  -0.132(z_{11})  -0.162(z_{12})  -0.131(z_{13})  -0.025(z_{14})\nonumber \\
&-0.144(z_{15}) -0.202(z_{16})  -0.161(z_{17}) + 0.435(z_{18}) -0.263 (z_{19}) + 0.221(z_{20}) -0.472(z_{21})\nonumber \\
& -0.141(z_{22}) -0.082(z_{23}) \nonumber \\
=& (3.82\times 10^{-4})W(\lambda4428) + (6.07\times 10^{-3})W(\lambda4964) - (5.74\times 10^{-2})W(\lambda 5494) + (1.08 \times 10^{-2}) W(\lambda 5513) \nonumber \\
&- (3.17 \times 10^{-2}) W(\lambda 5545) - (1.66 \times 10^{-2}) W(\lambda5546) + (3.77 \times 10^{-4}) W(\lambda 5769) - (2.84 \times 10^{-4}) W(\lambda 5780)\nonumber \\
&+ (3.06 \times 10^{-4}) W(\lambda 5797) + (3.50 \times 10^{-3}) W(\lambda5850) - (1.58 \times 10^{-2}) W(\lambda 6196) - (1.16 \times 10^{-2}) W(\lambda6270)\nonumber\\
&- (1.56 \times 10^{-3})W(\lambda 6284) - (3.42 \times 10^{-3}) W(\lambda6376) - (7.96 \times 10^{-3}) W(\lambda 6379) - (5.73 \times 10^{-3})W(\lambda6614)\nonumber\\
&- (1.22)E(B-V) + (4.74 \times 10^{-22}) N(H \textsc{i}) - (1.01 \times 10 ^{-21}) N(H_2) + (1.86 \times 10^{-22}) N(H)\nonumber\\
&- 2.02 f(H_2) + (6.60 \times 10^{-1}) F_{\star} + (5.22 \times 10^{-1}) W(\lambda5797)/W(\lambda5780) + 0.99 \nonumber
\end{align}

\begin{align}
\label{eqn:PC4_eqn}
  PC_4 =& 0.134(z_1) + 0.035(z_2)  -0.237(z_3) + 0.042(z_4)  -0.137(z_5)  -0.037(z_6) + 0.001(z_7)\\
  &+ 0.115(z_8) -0.230(z_9)  -0.145(z_{10}) -0.017(z_{11}) + 0.006(z_{12}) + 0.085(z_{13})  -0.169(z_{14}) \nonumber \\
  & -0.179(z_{15}) -0.030(z_{16}) + 0.200(z_{17}) + 0.238(z_{18}) + 0.109(z_{19}) + 0.230(z_{20}) + 0.126(z_{21}) \nonumber \\
  &+ 0.644(z_{22}) -0.394(z_{23}) \nonumber \\
=& (3.82\times 10^{-4})W(\lambda4428) + (6.07\times 10^{-3})W(\lambda4964) - (5.75\times 10^{-2})W(\lambda 5494) + (1.08 \times 10^{-2}) W(\lambda 5513) \nonumber \\
&- (3.17 \times 10^{-2}) W(\lambda 5545) - (1.66 \times 10^{-2}) W(\lambda5546) + (3.77 \times 10^{-4}) W(\lambda 5769) + (1.49 \times 10^{-3}) W(\lambda 5780)\nonumber \\
&- (8.27 \times 10^{-3}) W(\lambda 5797) - (1.07 \times 10^{-2}) W(\lambda5850) - (2.06 \times 10^{-3}) W(\lambda 6196) + (4.50 \times 10^{-4}) W(\lambda6270)\nonumber\\
&+ (1.01 \times 10^{-3})W(\lambda 6284) - (2.28 \times 10^{-2}) W(\lambda6376) - (9.92 \times 10^{-3}) W(\lambda 6379) - (8.51 \times 10^{-4})W(\lambda6614)\nonumber\\
& + 1.51 E(B-V) + (2.59 \times 10^{-22}) N(H \textsc{i}) + (4.19 \times 10 ^{-22}) N(H_2) + (1.93 \times 10^{-22}) N(H)\nonumber\\
&+ (5.37 \times 10^{-1}) f(H_2) + 3.01 F_{\star} - (2.51 \times 10^{-1}) W(\lambda5797)/W(\lambda5780) - 1.84 \nonumber
\end{align}

\section{Additional Figures}
As explained in the text and illustrated for 3 DIBs in
Fig.~\ref{fig:enviro}, we can use different normalizations to better
show the environmental responses of the DIBs. Below are similar
figures, but for the remaining DIBs.

\begin{figure}
\centering
{\includegraphics[width=17cm, trim={1.5cm, 2cm, 1.5cm, 2cm}]{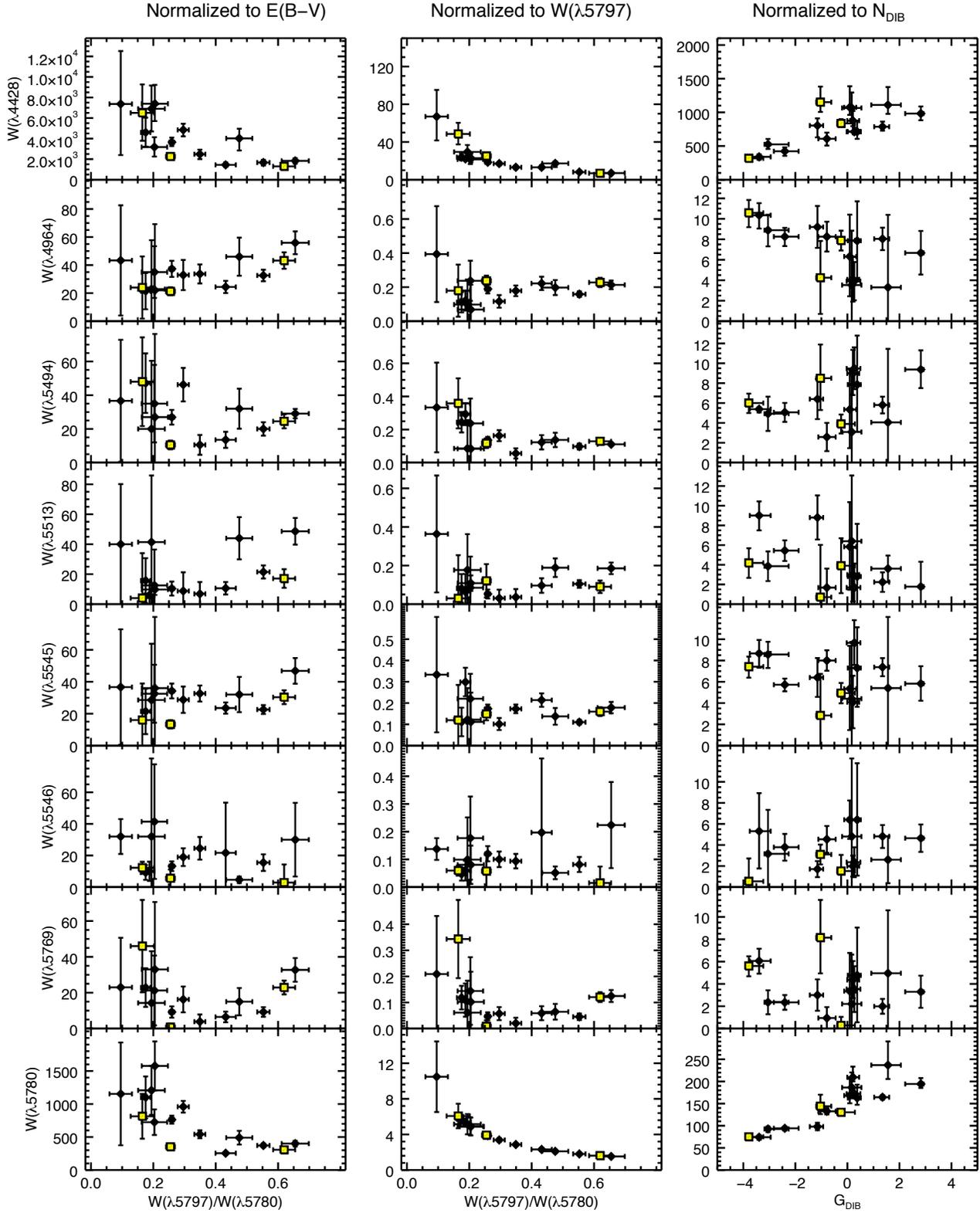}}
\caption{\label{fig:DIB_enviro} (\textit{Left}:) DIBs normalized to E(B-V) as a function of W($\lambda$5797)/W($\lambda$5780). (\textit{Centre}:) DIBs normalized to W($\lambda$5797) as a function of W($\lambda$5797)/W($\lambda$5780). (\textit{Right}:) DIBs normalized to N$_{\text{DIB}}$ as a function of G$_{\text{DIB}}$. Be stars are indicated by yellow squares. $\lambda$5513 for HD~40111, and $\lambda$5513 for both HD~15137 and HD~35149 have best value EW measurements equal to zero. Since fractional uncertainties are undefined, these data points have been excluded from the plots. HD~143275 has been excluded from the left plots because it has an E(B-V) value of zero.}
\end{figure}

\addtocounter{figure}{-1}

\begin{figure*}
\centering
{\includegraphics[width=17cm, trim={1.5cm, 2cm, 1.5cm, 2cm}]{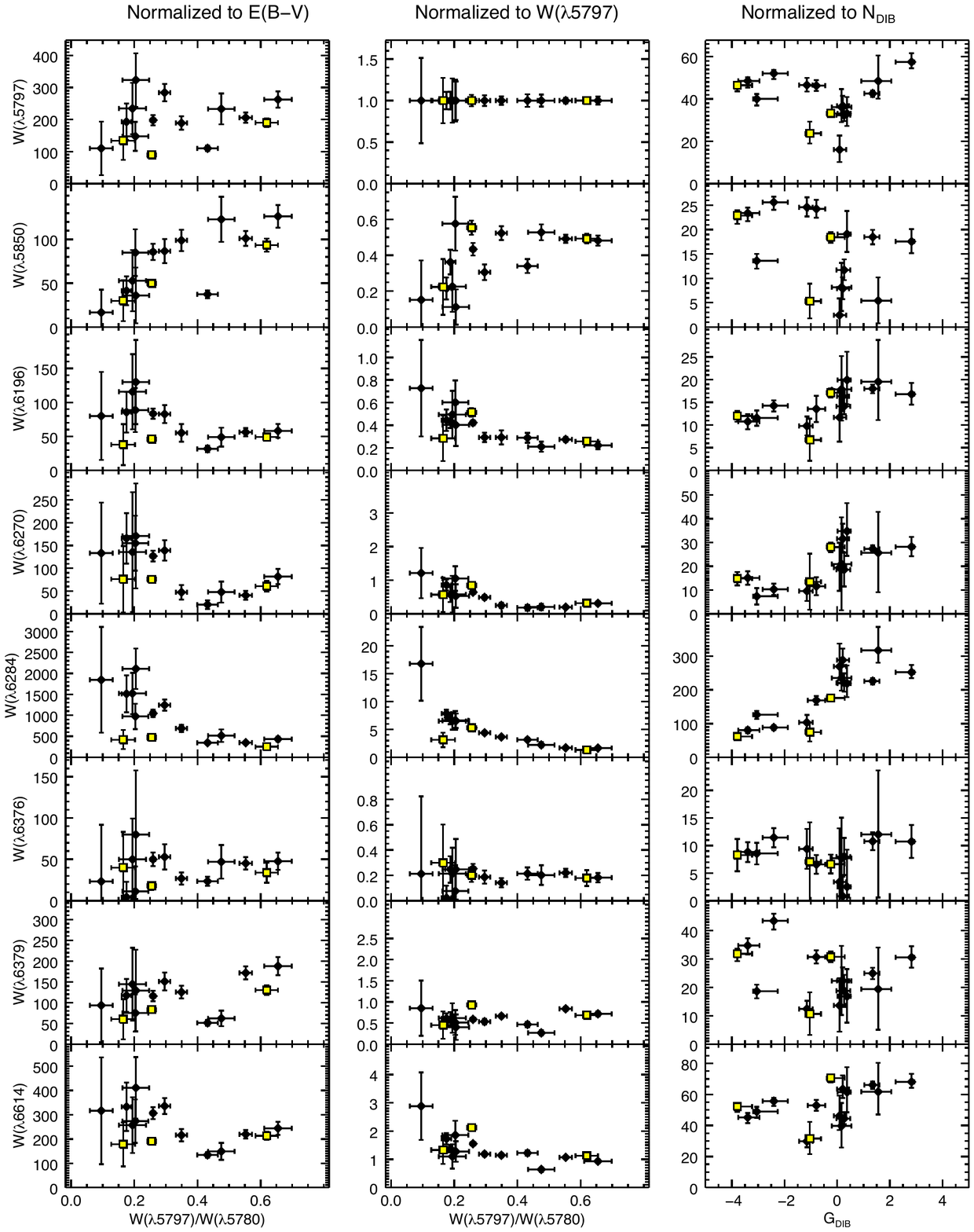}}
\caption{\label{fig:DIB_enviro} (Continued) (\textit{Left}:) DIBs normalized to E(B-V) as a function of W($\lambda$5797)/W($\lambda$5780). (\textit{Centre}:) DIBs normalized to W($\lambda$5797) as a function of W($\lambda$5797)/W($\lambda$5780). (\textit{Right}:) DIBs normalized to N$_{\text{DIB}}$ as a function of G$_{\text{DIB}}$. Be stars are indicated by yellow squares. HD~143275 has been excluded from the left plots because it has an E(B-V) value of zero.}
\end{figure*}

\clearpage
%\bibliographystyle{aasjournal}
%\bibliography{Biblio}

\end{document}

%% file: targettable.tex
\object{HD 15137}   &                   & 02 27 59.81  & +52 32 57.6    & 7.86 & 0.24 & 1.29$_{-0.40}^{+0.57}$ & 1.86$_{-0.12}^{+0.26}$ & 0.22$_{-0.06}^{+0.09}$ & 0.37$\pm$0.09  & 0.30$\pm$0.02 & -9.58  & 1  & ELODIE\\
\object{HD 22951}   &  40 Per           & 03 42 22.65  & +33 57 54.1    & 4.98 & 0.19 & 1.10$_{-0.32}^{+0.35}$ & 2.88$_{-0.98}^{+1.48}$ & 0.35$_{-0.18}^{+0.27}$ & 0.73$\pm$0.05  & 0.35$\pm$0.02 & 12.47  & 1  & ELODIE\\
\object{HD 23180}   &  o Per            & 03 44 19.13  & +32 17 17.7    & 3.86 & 0.22 & 0.76$_{-0.23}^{+0.26}$ & 3.98$_{-1.16}^{+1.64}$ & 0.51$_{-0.24}^{+0.33}$ & 0.84$\pm$0.06  & 0.65$\pm$0.04 & 13.45  & 2  & ELODIE\\
\object{HD 23630}   &  $\eta$ Tau       & 03 47 29.08  & +24 06 18.5    & 2.87 & 0.05 & 0.22$_{-0.07}^{+0.10}$ & 0.35$_{-0.12}^{+0.18}$ & 0.28$_{-0.15}^{+0.23}$ & 0.89$\pm$0.10  & 0.16$\pm$0.04 & 16.76  & 2  & ELODIE\\
\object{HD 24398}   &  $\zeta$ Per      & 03 54 07.92  & +31 53 01.1    & 2.88 & 0.27 & 0.63$_{-0.07}^{+0.06}$ & 4.68$_{-1.59}^{+2.40}$ & 0.59$_{-0.31}^{+0.46}$ & 0.88$\pm$0.05  & 0.55$\pm$0.02 & 14.54  & 2  & ELODIE\\
\object{HD 24534}   &  X Per            & 03 55 23.08  & +31 02 45.0    & 6.10 & 0.31 & 0.54$_{-0.07}^{+0.08}$ & 8.32$_{-0.73}^{+0.80}$ & 0.76$_{-0.11}^{+0.13}$ & 0.90$\pm$0.06  & 0.62$\pm$0.04 & 14.5  & 5  & ELODIE\\
\object{HD 24760}   &  $\epsilon$ Per   & 03 57 51.23  & +40 00 36.8    & 2.90 & 0.07 & 0.25$_{-0.05}^{+0.05}$ & 0.33$_{-0.15}^{+0.27}$ & 0.21$_{-0.14}^{+0.25}$ & 0.68$\pm$0.04  & 0.18$\pm$0.02 & 7.06  & 2  & ELODIE\\
\object{HD 24912}   &  $\xi$ Per        & 03 58 57.90  & +35 47 27.7    & 4.04 & 0.26 & 1.29$_{-0.24}^{+0.26}$ & 3.39$_{-0.99}^{+1.40}$ & 0.35$_{-0.15}^{+0.21}$ & 0.83$\pm$0.02  & 0.26$\pm$0.01 & 11.2  & 1  & ELODIE\\
\object{HD 27778}   &  62 Tau           & 04 23 59.76  & +24 18 03.6    & 6.33 & 0.34 & 0.22$_{-0.22}^{+0.55}$ & 5.25$_{-0.88}^{+1.06}$ & 0.82$_{-0.27}^{+0.45}$ & 1.19$\pm$0.07  & 0.43$\pm$0.03 & 15.22  & 2  & ELODIE\\
\object{HD 35149}   &  23 Ori           & 05 22 50.00  & +03 32 40.0    & 5.00 & 0.08 & 0.43$_{-0.13}^{+0.12}$ & 0.03$_{-0.03}^{+0.00}$ & 0.02$_{-0.02}^{+0.00}$ & 0.54$\pm$0.11  & 0.20$\pm$0.04 & 24.09  & 2  & UVES\\
\object{HD 35715}   &  $\Psi$ Ori       & 05 26 50.23  & +03 05 44.4    & 4.60 & 0.03 & 0.31$_{-0.13}^{+0.13}$ & 6$\pm 2\times 10^{-6}${} & 4$\pm 2\times 10^{-6}$ & 0.66$\pm$0.11  & 0.10$\pm$0.04 & 25.2  & 1  & ELODIE\\
\object{HD 36822}   &  $\varphi^1$ Ori     & 05 34 49.24  & +09 29 22.5    & 4.40 & 0.07 & 0.65$_{-0.12}^{+0.13}$ & 0.21$_{-0.06}^{+0.09}$ & 0.06$_{-0.03}^{+0.04}$ & 0.74$\pm$0.08  & 0.19$\pm$0.04 & 25.53  & 1  & ELODIE\\
\object{HD 36861}   &  $\lambda$ Ori A  & 05 35 08.28  & +09 56 03.0    & 3.30 & 0.10 & 0.60$_{-0.16}^{+0.16}$ & 0.13$_{-0.05}^{+0.08}$ & 0.04$_{-0.02}^{+0.04}$ & 0.57$\pm$0.04  & 0.48$\pm$0.04 & 25.2  & 3  & ELODIE\\
\object{HD 40111}   &  139 Tau          & 05 57 59.66  & +25 57 14.1    & 4.82 & 0.10 & 0.79$_{-0.15}^{+0.16}$ & 0.54$_{-0.20}^{+0.31}$ & 0.12$_{-0.07}^{+0.10}$ & 0.49$\pm$0.04  & 0.20$\pm$0.04 & 15.29  & 2  & ELODIE\\
\object{HD 110432}  &  BZ Cru           & 12 42 50.27  & $-$63 03 31.0  & 5.32 & 0.39 & 0.71$_{-0.21}^{+0.29}$ & 4.37$_{-0.38}^{+0.42}$ & 0.55$_{-0.11}^{+0.13}$ & 1.17$\pm$0.11  & 0.25$\pm$0.01 & 6.8  & 3  & UVES\\
\object{HD 143275}  &  $\delta$ Sco     & 16 00 20.01  & $-$22 37 18.1  & 2.29 & 0.00 & 1.41$_{-0.29}^{+0.29}$ & 0.26$_{-0.09}^{+0.15}$ & 0.03$_{-0.02}^{+0.03}$ & 0.90$\pm$0.03  & 0.19$\pm$0.02 & -10.90  & 2  & UVES\\
\object{HD 144217}  &  $\beta^1$ Sco    & 16 05 26.23  & $-$19 48 19.6  & 2.62 & 0.18 & 1.23$_{-0.11}^{+0.12}$ & 0.68$_{-0.09}^{+0.10}$ & 0.10$_{-0.02}^{+0.02}$ & 0.81$\pm$0.02  & 0.11$\pm$0.01 & -8.95  & 2  & UVES\\
\object{HD 145502}  &  $\nu$ Sco        & 16 11 59.74  & $-$19 27 38.5  & 4.13 & 0.20 & 1.17$_{-0.59}^{+0.56}$ & 0.78$_{-0.23}^{+0.32}$ & 0.12$_{-0.07}^{+0.08}$ & 0.80$\pm$0.11  & 0.18$\pm$0.01 & -8.49  & 2  & ELODIE\\
\object{HD 147165}  &  $\sigma$ Sco     & 16 21 11.32  & $-$25 35 34.0  & 2.91 & 0.31 & 2.19$_{-0.87}^{+0.90}$ & 0.62$_{-0.18}^{+0.25}$ & 0.05$_{-0.03}^{+0.04}$ & 0.76$\pm$0.06  & 0.13$\pm$0.01 & -6.26  & 2  & UVES\\
\object{HD 147933}  &  $\rho$ Oph A     & 16 25 35.10  & $-$23 26 48.7  & 5.02 & 0.37 & 4.27$_{-0.80}^{+0.98}$ & 3.72$_{-1.09}^{+1.53}$ & 0.15$_{-0.07}^{+0.09}$ & 1.09$\pm$0.08  & 0.27$\pm$0.03 & -8.02  & 2  & UVES\\
\object{HD 149757}  &  $\zeta$ Oph      & 16 37 09.54  & $-$10 34 01.5  & 2.58 & 0.29 & 0.52$_{-0.04}^{+0.02}$ & 4.47$_{-0.75}^{+0.90}$ & 0.63$_{-0.17}^{+0.20}$ & 1.05$\pm$0.02  & 0.50$\pm$0.04 & -14.98  & 2  & UVES\\
\object{HD 164284}  &  66 Oph           & 18 00 15.80  & +04 22 07.0    & 4.78 & 0.11 & 0.42$_{-0.39}^{+0.23}$ & 0.71$_{-0.21}^{+0.29}$ & 0.25$_{-0.20}^{+0.18}$ & 0.89$\pm$0.18  & 0.15$\pm$0.02 & -15.32  & 1  & ELODIE\\
\object{HD 170740}  &                   & 18 31 25.69  & $-$10 47 45.0  & 5.76 & 0.38 & 1.07$_{-0.47}^{+0.59}$ & 7.24$_{-1.22}^{+1.47}$ & 0.58$_{-0.18}^{+0.22}$ & 1.02$\pm$0.11  & 0.26$\pm$0.01 & -12.9  & 6  & UVES\\
\object{HD 198478}  &  55 Cyg           & 20 48 56.29  & +46 06 50.9    & 4.86 & 0.43 & 2.04$_{-0.63}^{+0.84}$ & 7.41$_{-2.17}^{+3.06}$ & 0.42$_{-0.20}^{+0.27}$ & 0.81$\pm$0.05  & 0.24$\pm$0.01 & -10.04  & 2  & ELODIE\\
\object{HD 202904}  &  $\upsilon$ Cyg   & 21 17 55.08  & +34 53 48.8    & 4.43 & 0.09 & 0.23$_{-0.23}^{+0.21}$ & 0.14$_{-0.05}^{+0.07}$ & 0.11$_{-0.10}^{+0.12}$ & 0.39$\pm$0.11  & 0.13$\pm$0.05 & -12.90 & 4  & ELODIE\\
\object{HD 207198}  &                   & 21 44 53.28  & +62 27 38.0    & 5.96 & 0.47 & 3.39$_{-0.50}^{+0.59}$ & 6.76$_{-0.59}^{+0.65}$ & 0.28$_{-0.05}^{+0.05}$ & 0.90$\pm$0.03  & 0.53$\pm$0.01 & -15.28  & 2  & ELODIE\\
\object{HD 209975}  &  19 Cep           & 22 05 08.79  & +62 16 47.3    & 5.11 & 0.27 & 1.29$_{-0.38}^{+0.41}$ & 1.20$_{-0.41}^{+0.62}$ & 0.16$_{-0.09}^{+0.12}$ & 0.57$\pm$0.26  & 0.31$\pm$0.01 & -11.39  & 2  & ELODIE\\
\object{HD 214680}  &  10 Lac           & 22 39 15.68  & +39 03 01.0    & 4.88 & 0.08 & 0.50$_{-0.15}^{+0.14}$ & 0.17$_{-0.04}^{+0.05}$ & 0.06$_{-0.03}^{+0.03}$ & 0.50$\pm$0.06  & 0.34$\pm$0.02 & -9.2 & 1  & ELODIE\\
\object{HD 214993}  &  12 Lac           & 22 41 28.65  & +40 13 31.6    & 5.23 & 0.06 & 0.58$_{-0.18}^{+0.20}$ & 0.43$_{-0.14}^{+0.22}$ & 0.13$_{-0.07}^{+0.10}$ & 0.68$\pm$0.10  & 0.17$\pm$0.02 & -9.44  & 1  & ELODIE\\
\object{HD 218376}  &  1 Cas            & 23 06 36.82  & +59 25 11.1    & 4.84 & 0.16 & 0.89$_{-0.26}^{+0.28}$ & 1.41$_{-0.48}^{+0.73}$ & 0.24$_{-0.13}^{+0.19}$ & 0.60$\pm$0.06  & 0.28$\pm$0.01 & -12.65  & 1  & ELODIE\\